\documentclass[preprint2,twocolumn,times,tighten]{aastex63}

\usepackage{graphicx,times}
\usepackage{subfigure}
\newcommand{\be}{\begin{equation}}
\usepackage{threeparttable}
\usepackage{booktabs}
\newcommand{\ee}{\end{equation}}
\newcommand{\bea}{\begin{eqnarray}}
\newcommand{\eea}{\end{eqnarray}}
\newcommand*{\sy}{\color{black}}
\newcommand*{\xu}{\color{black}}

\usepackage{amsmath}
\usepackage{cases}
\usepackage{longtable}
\usepackage{hyperref}
\usepackage{epstopdf}
\usepackage{amsmath,bm}
\usepackage{amssymb}
\usepackage{natbib}
\usepackage{morefloats}
\usepackage{multirow}
\usepackage{array}
\usepackage{verbatim}
\usepackage{color}
\usepackage{lineno}
\begin{document}

\title{Shock acceleration with oblique and turbulent magnetic fields}


\author[0000-0002-0458-7828]{Siyao Xu}
\affiliation{Institute for Advanced Study, 1 Einstein Drive, Princeton, NJ 08540, USA; sxu@ias.edu
\footnote{Hubble Fellow}}


\author{Alex Lazarian}
\affiliation{Department of Astronomy, University of Wisconsin, 475 North Charter Street, Madison, WI 53706, USA; 
lazarian@astro.wisc.edu}

\affiliation{Centro de Investigación en Astronomía, Universidad Bernardo O’Higgins, Santiago, General Gana 1760, 8370993,Chile}

\begin{abstract}

We investigate shock acceleration in a realistic astrophysical environment with density inhomogeneities. 
The turbulence induced by the interaction of the shock precursor with upstream density fluctuations
amplifies both upstream and downstream magnetic fields via the turbulent dynamo. 
The dynamo-amplified turbulent magnetic fields 
(a) introduce variations of shock obliquities along the shock face,
(b) enable energy gain through a combination of shock drift and diffusive processes, 
(c) give rise to various spectral indices of accelerated particles, 
(d) regulate the diffusion of particles both parallel and perpendicular to the magnetic field, and 
(e) increase the shock acceleration efficiency.
Our results demonstrate that upstream density inhomogeneities and dynamo amplification of magnetic fields play 
an important role in shock acceleration, and thus shock acceleration depends on the condition of the ambient interstellar environment.  
The implications on understanding radio spectra of supernova remnants are also discussed. 
\end{abstract}


\section{Introduction}

Shock acceleration is a fundamental particle acceleration mechanism in both 
space physics and astrophysics
\citep{Uro19,Mar20,Liu21}.
As the most accepted model for shock acceleration, the diffusive shock acceleration (DSA) mechanism for a parallel shock, 
where the shock normal is parallel to the upstream magnetic field,
has been extensively discussed in the literature 
\citep{Axf77,Kry77,Bell78,Blan78,Drury83,Jon91,Longairbook,Ach00}.
Particles are diffusively scattered by magnetic fluctuations in both upstream and downstream regions and 
repeatedly accelerated in multiple crossings of a parallel shock. 
DSA of a parallel shock predicts a universal spectral index of accelerated particles determined by the shock compression ratio. 
However, radio and gamma-ray observations of supernova remnants (SNRs) suggest a variety of particle spectral indices 
\citep{Oni13,Ack13,Uro14}.

In a more general case with an oblique shock,
shock drift acceleration (SDA) takes place. 
Particles drift along the shock front due to the the change in magnetic field
and get accelerated by the convective electric field parallel to the drift motion
\citep{Son69,Wu84,Arm85,BaMel01}.
SDA is believed to be important for particle acceleration 
at bow shocks and interplanetary shocks 
\citep{Arm76,Lop89}
and a pre-acceleration mechanism for DSA
\citep{Caps14,vanM18}.
In the presence of magnetic fluctuations, a more general model that incorporates both SDA and DSA 
has been proposed 
\citep{Jok82,DecV86,Deck88,Ostr88,Jok87,Kirk89}.
The diffusive and drift processes simultaneously operate for particles to experience multiple shock encounters
and drift acceleration at each encounter. 
The resulting energy spectrum of accelerated particles and 
acceleration rate are found to be significantly different from those of a 
parallel shock 
\citep{Jok87,Kirk89,NaTa95,Han19,Be21}.


{\xu Shock acceleration} 
is traditionally formulated under the assumption of a uniform upstream density distribution, 
{\xu and the generation of upstream magnetic fluctuations appeals to 
plasma streaming instabilities driven by the cosmic ray (CR) precursor 
\citep{Bel13}.}
Turbulence and density fluctuations are ubiquitous in the interstellar medium (ISM). 
Observations reveal a large variety of turbulent density spectra in different interstellar phases 
\citep{Armstrong95,CheL10,Laz09rev,Hen12,XuZ16,XuZ17,Xup20}.
For a supernova (SN) shock propagating through the ISM, the interaction of the CR precursor 
with upstream density inhomogeneities induces vorticity and turbulence, 
and the turbulence amplifies the preshock and postshock magnetic fields via the turbulent dynamo 
\citep{BJL09,Ino09,Dru12,Del16,XuL17,Xupar19}.
The dynamo-amplified turbulent magnetic fields introduce 
variations of shock obliquities and 
affect diffusion of particles both parallel and perpendicular to magnetic fields. 
Local oblique shocks are created in the presence of turbulent magnetic fields 
irrespective of the initial obliquity angle between the shock normal and the interstellar magnetic field. 
Based on the modern understanding of magnetohydrodynamic (MHD) turbulence
\citep{GS95,LV99},
both parallel and perpendicular diffusion of particles 
strongly depends on the properties of MHD turbulence
\citep{YL08,XY13,LY13,LX21}.
As both shock obliquity and particle diffusion are important ingredients of shock acceleration, it is necessary to examine 
their effects on particle energy spectrum and acceleration efficiency 
in the presence of dynamo-amplified turbulent magnetic fields.

In this work, we will investigate the shock acceleration in a realistic situation 
by taking into account upstream density inhomogeneities and dynamo amplification of magnetic fields. 
We consider the shock acceleration of relativistic particles with the particle speed much larger than the shock speed 
and the gyroradius much larger than the shock thickness. 
In Section 2, we first briefly review the basic physics of DSA of a parallel shock. 
The more general case of an oblique shock involving both DSA and SDA is discussed in Section 3. 
In Section 4, we focus on the shock acceleration in the presence of dynamo-amplified turbulent magnetic fields. 
The corresponding acceleration time in different scenarios is studied in Section 5. 
Implications of our results on particle energy spectra of SNRs are discussed in Section 6. 
Conclusions are presented in Section 7.

 \section{DSA at parallel shocks}
 \label{sec: paralsh}
 
We first briefly review the theory of DSA 
for a strong parallel shock propagating through a homogeneous diffuse medium. 
Here we do not consider the nonlinear modification of shock structure by the CR pressure
\citep{Rey08}.
Under the assumption of efficient scattering, 
particles are coupled to the fluid and have isotropic velocity distribution 
on both sides of the shock
\citep{Park65,Skil75}.
The scattering causes particles to repeatedly diffuse across the shock front
and undergo the same head-on collisions with 
scattering centers in the upstream and downstream flows. 

As formulated by
\citet{Bell78},
in one cycle of diffusion from upstream to downstream and back,
the average fractional energy increase of relativistic particles is 
\begin{equation}
    \frac{\Delta E}{E_0}  = \frac{4}{3} \frac{V}{c},
\end{equation} 
where $c$ is the light speed, $V = 3/4 U_{sh}$, 
$U_{sh}$ is the shock speed, 
and the shock compression ratio is $4$ with the ratio of specific heats $5/3$.
This is a small increase in energy for $U_{sh}\ll c$.
Then the ratio of particle energy $E$ to its initial value $E_0$ is 
\begin{equation}\label{eq: pasbe}
   \epsilon = \frac{E}{E_0} = \frac{\Delta E + E_0}{ E_0} = 1+ \frac{4}{3} \frac{V}{c}  = 1 + \frac{U_\text{sh}}{c}.
\end{equation} 
The mean probability that the particle returns to the shock per cycle is 
\begin{equation}
   P = 1 - \frac{U_{sh}}{c},
\end{equation} 
with a small probability $U_{sh}/c$
that the particle is advected away from the shock. 
Given $U_{sh}\ll c$, 
the competition between the energy gain and escape results in a power-law 
particle energy spectrum, 
\begin{equation}\label{eq: pwspec}
  N(E) \propto E^{-1 + (\ln P / \ln \epsilon)} = E^{-2}.
\end{equation} 
Its energy spectral index is uniquely determined by the shock compression.

\section{Oblique shocks involving both DSA and SDA}
\label{sec:oblshock}
 
Oblique shocks are commonly seen.
For a spherical shock expanding into a large-scale uniform magnetic field, the average shock obliquity is $60^\circ$. 
With preshock turbulent magnetic fields, locally oblique shocks are created regardless of the initial shock obliquity.

At an oblique shock with a subluminal shock front, 
{\xu i.e., $U_{sh}/\cos \alpha_1 <c$, where $\alpha_1$ is the angle between the shock normal and the upstream magnetic field,}
a particle incident from upstream
can be either reflected from the shock front by the mirror force of compressed magnetic field 
or transmitted into the downstream region. 
Due to the shock compression of magnetic field, 
a particle undergoes a gradient $B$ drift along the convection electric field, 
{\xu which is $-{\bf U_{sh}}\times {\bf B}/c$
and parallel to 
the shock surface in the shock frame.}
The acceleration via the drift mechanism during a single shock encounter is known as SDA
\citep{Sarr74,Arms77}.
A large fractional energy increase for reflected particles can be achieved 
in a narrow range of initial pitch angles and at large obliquities
\citep{Deck88,kirkbook94,BaMel01}.

In the presence of turbulent magnetic fields, 
the spatial diffusion of particles due to pitch angle scattering or tangling of field lines (see Section \ref{ssec: turdyn})
in both upstream and downstream regions enables 
multiple encounters between particles and the shock front.
Thus both SDA and DSA contribute to particle acceleration at an oblique shock
\citep{Jok82,Deck88,Kirk89}.

\subsection{Spectral index of accelerated particles} 
 
Under the assumptions of isotropic particle distribution in the local fluid frames
and conservation of magnetic moment during the drift acceleration, 
the particle acceleration process involving both SDA and DSA at oblique shock fronts 
has been earlier studied 
(e.g., \citealt{Drury83,Ostr88,NaTa95}).
Isotropic particle distribution requires efficient isotropic scattering.  
For the second assumption, 
despite the discontinuous change in magnetic field strength at the shock, 
the adiabatic approximation is valid when the drift motion is slower than the gyromotion, 
{\xu that is, a particle's orbit intersects the shock front many times}  
\citep{kirkbook94}, 
as seen in test particle simulations 
\citep{Deck88}.
It also requires $r_g<\lambda_\|$ for the trajectories of 
particles interacting with the shock front to be unperturbed, 
where $r_g$ is the particle gyroradius, and $\lambda_\|$ is the scattering mean free path.


(1) Limit case with {\xu small $\beta_1$}.

When {\xu $V_1 \ll c$}, 
the approximate mean fractional energy increase per cycle is 
\begin{equation}\label{eq: orlbd}
\begin{aligned}
    d &= \frac{\Delta E}{E_0} 
        \approx  \frac{V_1}{c} = \beta_1,
\end{aligned}
\end{equation} 
where 
\begin{equation}
   V_1 = \frac{U_1}{\cos \alpha_1},  ~~ V_2 = \frac{U_2}{\cos\alpha_2},
\end{equation}
$U_i$ is the shock speed in the local fluid frame
\citep{hud65},
$\alpha_i$ is the angle between the shock normal and the magnetic field, and 
subscripts $i =1$, $2$ refer to the upstream and downstream regions.  
This result can be derived from the calculations by, e.g., \citet{Drury83,Ostr88}, 
and their calculations are carried out up to the first order in $\beta_1$.
Accordingly, we have  
\begin{equation}\label{eq: oblbeta}
   \epsilon = 1+d 
            \approx 1+ \beta_1.
\end{equation} 
When $\beta_1 \ll 1$,
despite the complications arising from SDA, 
the acceleration at an oblique shock has a simple form of $\epsilon$ similar to that (Eq. \eqref{eq: pasbe}) for a parallel shock.
To the same approximation order $O(\beta_1)$, 
the probability that a particle escapes downstream is 
\citep{Bell78,Drury83}
\begin{equation}\label{eq: orlbpesp}
   P_\text{esp} \approx (1-\mu_0^2) 4 \beta_2 = \beta_1, 
\end{equation}
and thus 
\begin{equation}\label{eq: oblres}
  P  = 1 - P_\text{esp} \approx  
   1 - \beta_1,
\end{equation} 
where $\beta_2 = V_2 /c$,
$1-\mu_0^2$ is the approximate probability of transmission from upstream to downstream regions,  
\begin{equation}\label{eq: mu0bb}
   \mu_0 = \Big(1-\frac{B_1}{B_2}\Big)^\frac{1}{2},
\end{equation}
and $B_i$ is the magnetic field strength. We note that $V_1$ and $V_2$ are related by 
\begin{equation}
   \frac{V_1}{V_2} 
    = \frac{U_1}{U_2} (1-\mu_0^2)
    = 4 (1-\mu_0^2).
\end{equation} 
By combining Eqs. \eqref{eq: oblbeta} and \eqref{eq: oblres}, we see that 
\begin{equation}\label{eq: orlbspec}
\begin{aligned}
   \frac{\ln P}{\ln \epsilon} 
   & = \frac{\ln (1-\beta_1)}{\ln ( 1+ \beta_1)}.
\end{aligned}
\end{equation}
As $\beta_1 \ll 1$, there is 
\begin{equation}
    \frac{\ln P}{\ln \epsilon} \approx 
     -1.
\end{equation} 
The spectral index of $N(E)$ (see Eq. \eqref{eq: pwspec}) is the same as that for a parallel shock 
\citep{Bell78,Drury83}.

(2) General case.

The above result is only valid in the limit case with $\beta_1 \ll 1$.
When $V_1$ is a significant fraction of $c$, we follow the same approach in 
\citet{Ostr88}, 
but consider a general case without taking the approximation of a small $\beta_1$. 
The mean fractional energy increase per cycle is 
\begin{equation}\label{eq: gendos}
   d = P_{12} (d_{12}+ d_{21}) + P_r d_r, 
\end{equation}
where $P_{12}$ and $P_r$ are the probabilities of transmission and reflection, 
and $d_{12}$, $d_{21}$, and $d_r$ are the  mean fractional energy gain during transmission from upstream to downstream, 
transmission from downstream to upstream, and reflection. 
The probability of a particle returning to the shock per cycle is 
\begin{equation}
    P = 1 - P_\text{esp} = 1 - P_{12} \frac{4\beta_2}{(1+\beta_2)^2}.
\end{equation}
The derivation and detailed expressions of $d$ and $P$ are provided in Appendix \ref{sec:app}.

\begin{figure*}[ht]
\centering
\subfigure[]{
   \includegraphics[width=8.7cm]{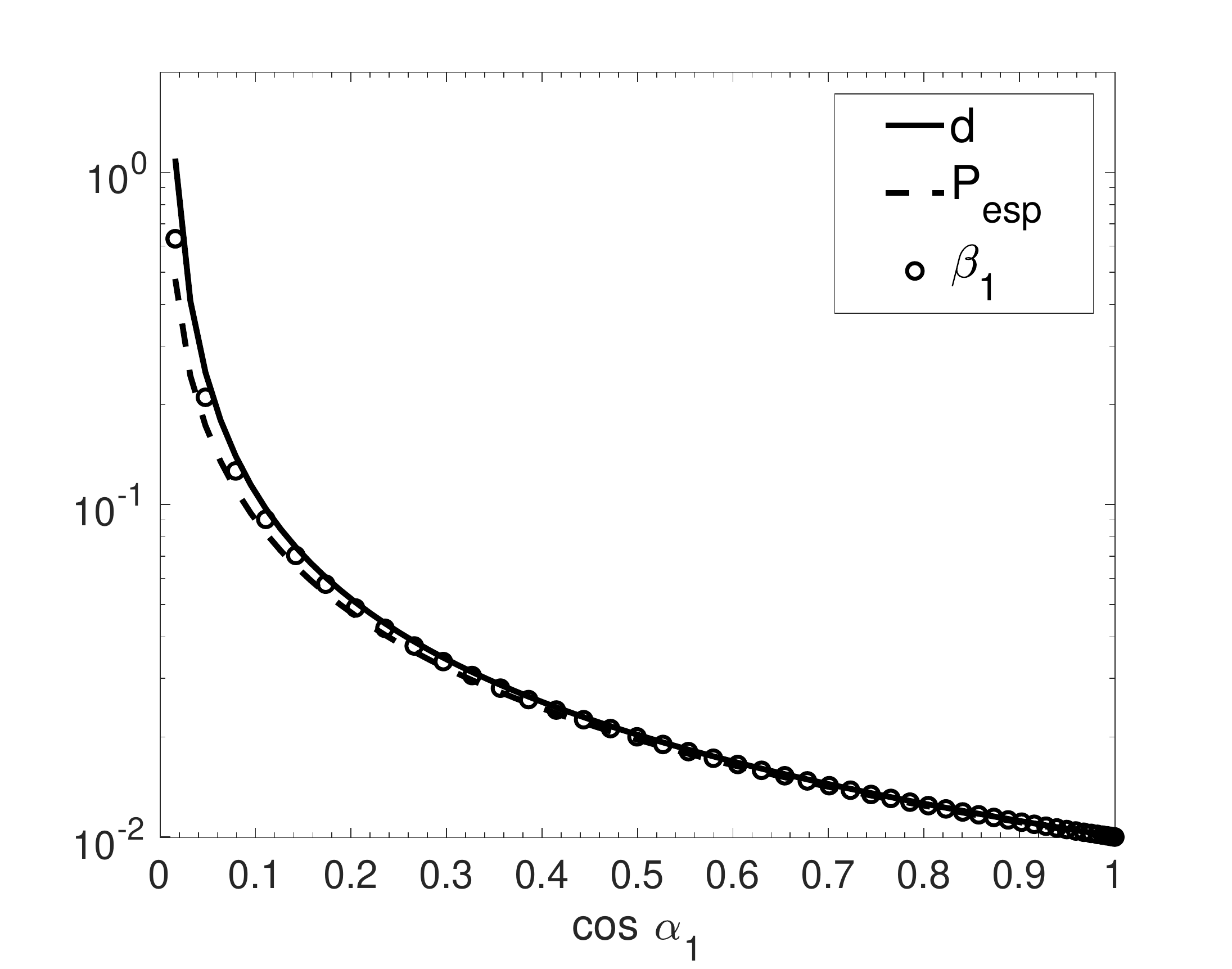}\label{fig: dpesp}}
\subfigure[]{
   \includegraphics[width=8.7cm]{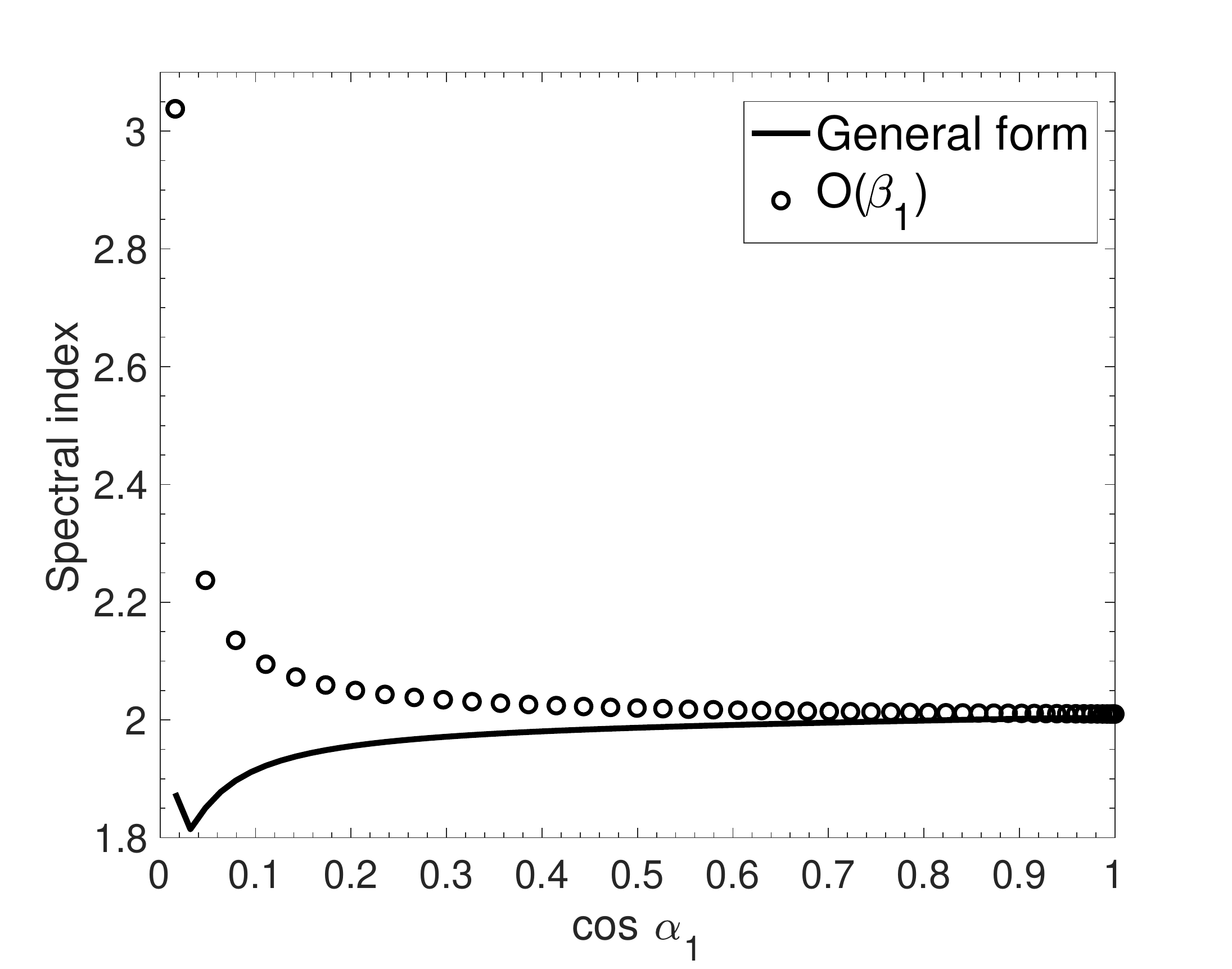}\label{fig: speind}}
\caption{$d$ and $P_\text{esp}$ in (a) and the energy spectral index of accelerated particles in (b) 
as a function of $\cos \alpha_1$ at $U_1=0.01c$ for an oblique shock. Solid and dashed lines correspond to the general solutions. 
Circles represent the approximate results in the limit case with a small $\beta_1$.
}
\label{fig: shoiso}
\end{figure*}

\begin{figure*}[ht]
\centering
\subfigure[]{
   \includegraphics[width=8.7cm]{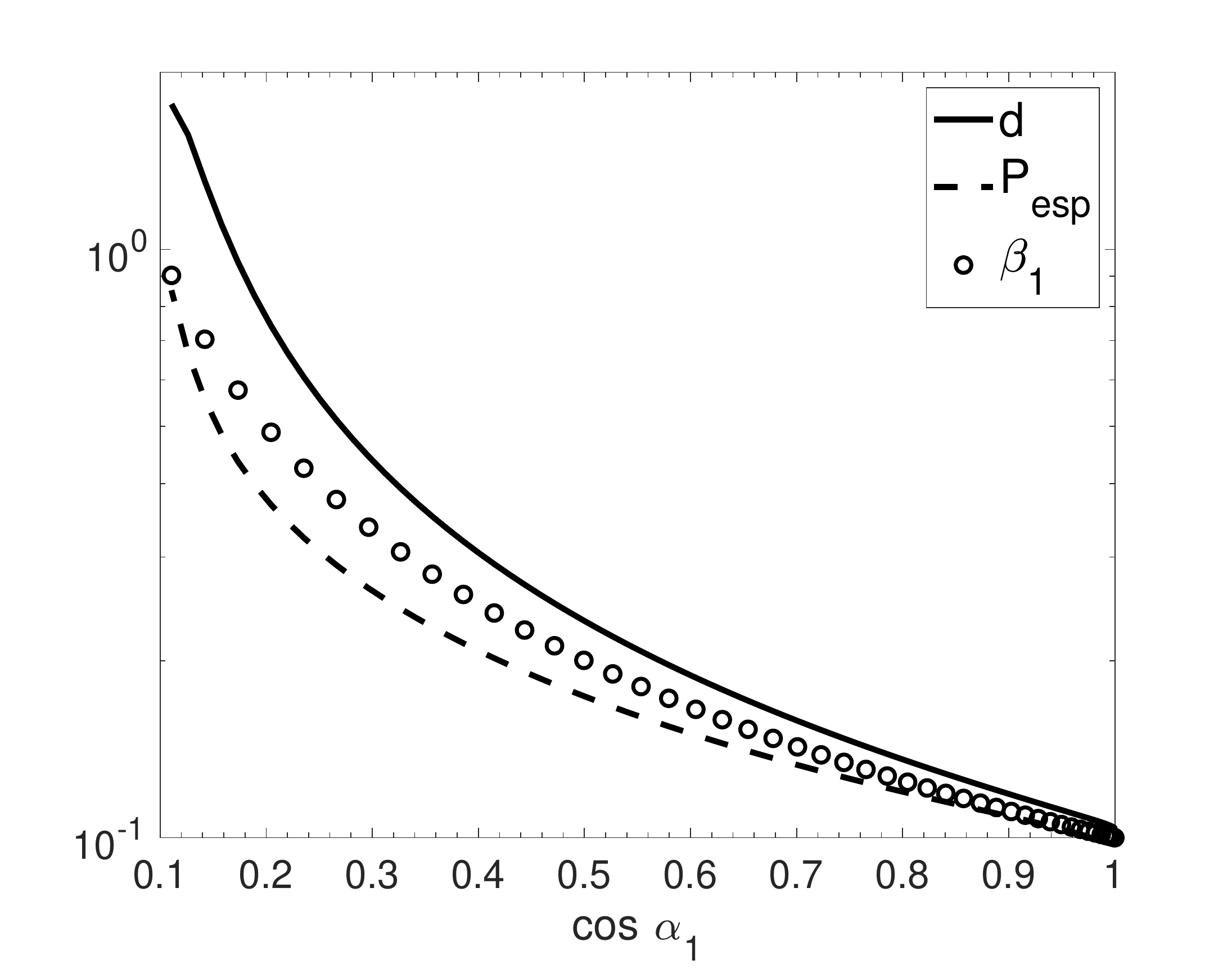}\label{fig: dpesp2}}
\subfigure[]{
   \includegraphics[width=8.7cm]{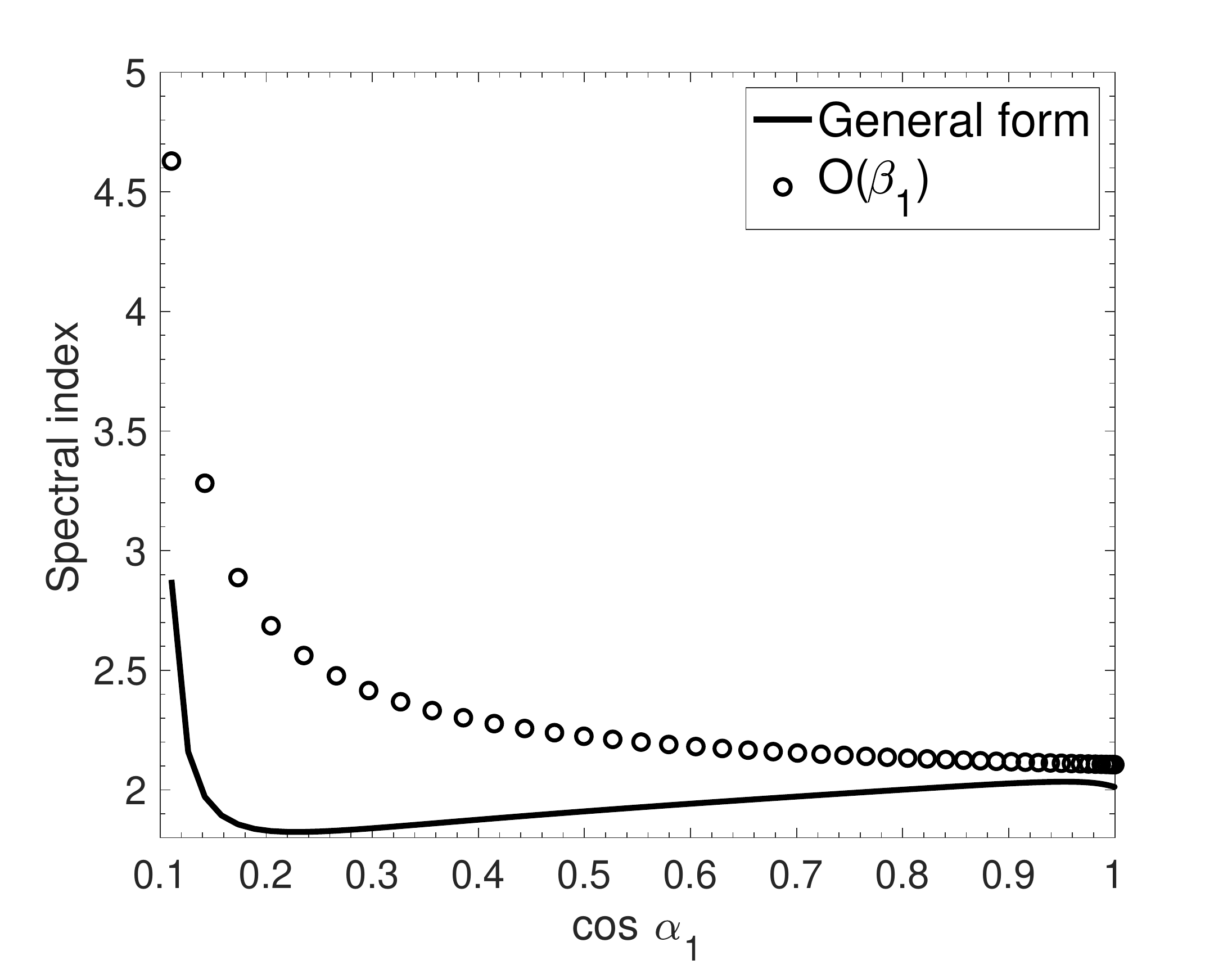}\label{fig: speind2}}
\caption{Same as Fig. \ref{fig: shoiso}, but for $U_1=0.1c$.}
\label{fig: shoiso2}
\end{figure*}

The numerically calculated results are presented in Fig. \ref{fig: shoiso} at $U_1 = 0.01 c$ and 
in Fig. \ref{fig: shoiso2} at $U_1 = 0.1 c$.
We also present the approximate results from Eqs. \eqref{eq: orlbd}, \eqref{eq: orlbpesp}, and \eqref{eq: orlbspec} as a comparison, 
but note that they are only valid when $\beta_1$ is small. 
We see that both $d$ and $P_\text{esp}$ increase with obliquity. 
At $U_1 = 0.01 c$, $\beta_1$ is small for most shock obliquities. 
The approximation $d \approx P_\text{esp} \approx \beta_1$ and the corresponding spectral index $\approx 2$
give a satisfactory agreement with the general solution
except for large shock obliquities.
As the general expression of $P_\text{esp}$
leads to a smaller value than $\beta_1$ toward large obliquities, 
the spectrum gradually becomes shallower with increasing obliquity, having the spectral index smaller than $2$. 
At $U_1 = 0.1 c$ (Fig. \ref{fig: shoiso2}), 
the approximation $d \approx P_\text{esp} \approx \beta_1$ and the spectral index $\approx 2$
can only stand at small obliquities for quasi-parallel shocks. 
We see $d > \beta_1$ and $P_\text{esp}<\beta_1$.
In particular, very large energy gain per cycle with $d>1$ is achieved at large obliquities. 
The spectral index gradually decreases with increasing obliquity. 
After reaching the minimum index $\approx 1.8$,
the spectrum becomes very steep due to the rapid rise of $P_\text{esp}$ at large obliquities.

(3) General case with modified energy gain.
 
\begin{figure*}[ht]
\centering
\subfigure[]{
   \includegraphics[width=8.7cm]{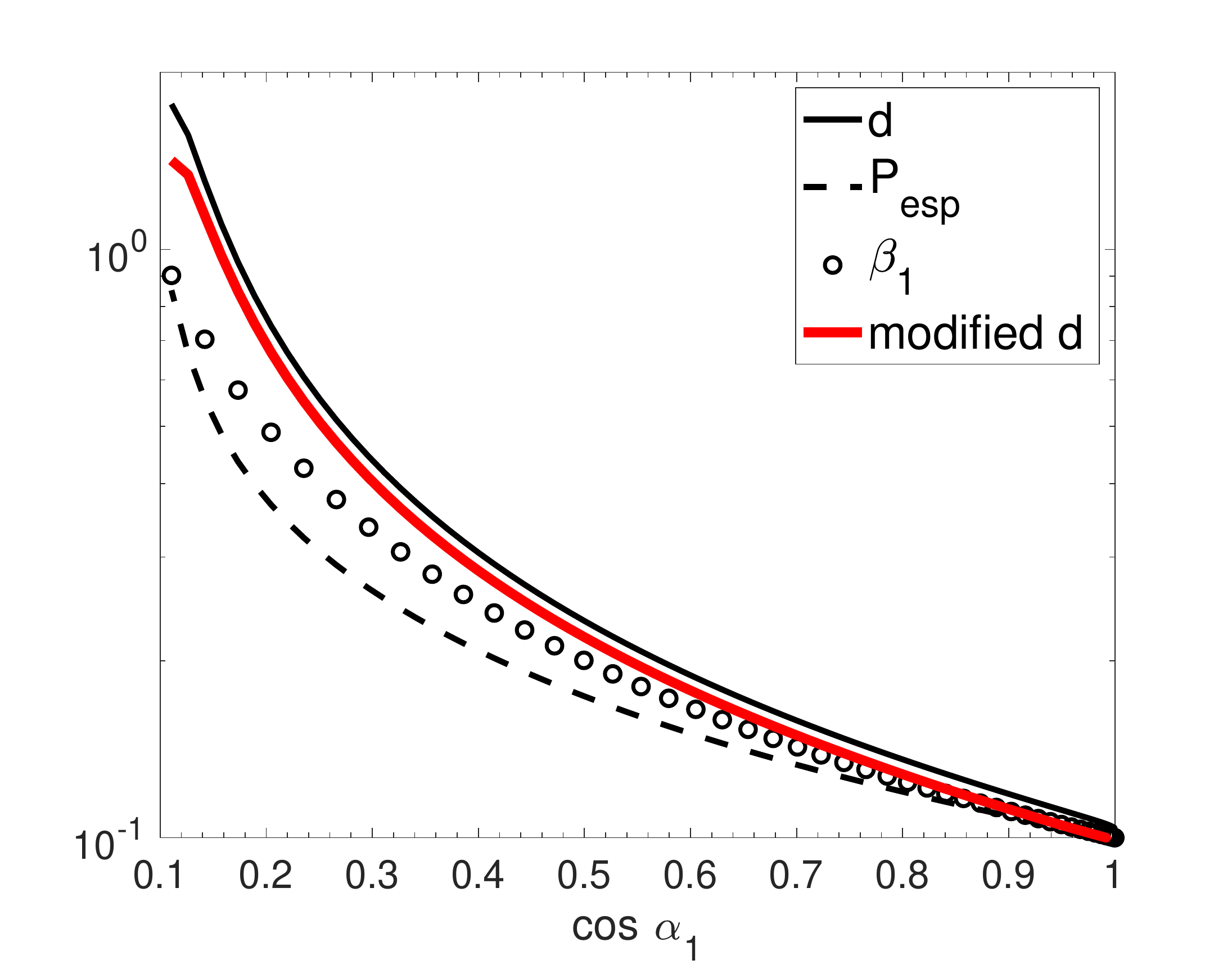}\label{fig: dpesp2mod}}
\subfigure[]{
   \includegraphics[width=8.7cm]{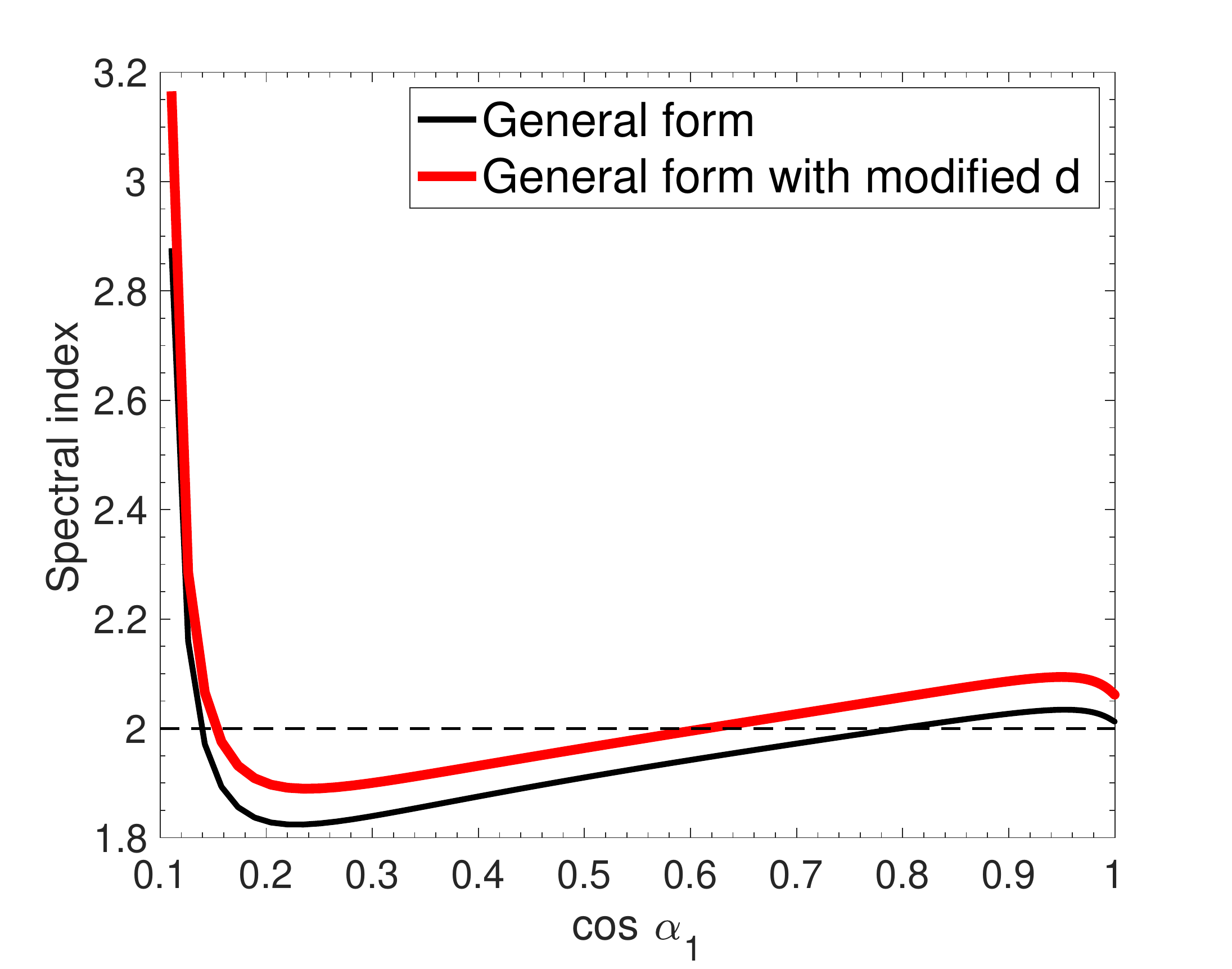}\label{fig: speind2mod}}
\caption{Same as Fig. \ref{fig: shoiso2}, but with the results using modified $d$ added (thick red lines).
The horizontal dashed line in (b) indicates the index value equal to $2$.}
\label{fig: shoiso2mod}
\end{figure*}

We note that in the above calculations on $d$, the escape of particles 
after transmission downstream
is not considered. 
With the escaping particles 
taken into account, $d$ should be modified as 
\begin{equation}\label{eq: gendoblo}
   d = P_{12}\bigg[1- \frac{4 \beta_2}{(1+\beta_2)^2}\bigg] (d_{12}+ d_{21}) 
         + P_\text{esp} d_{12}
         + P_r d_r.
\end{equation}
At $U_1 = 0.01 c$, $P_\text{esp}$ is small, and thus the above modification is insignificant. 
At $U_1 = 0.1 c$, the modified $d$ is smaller than that in Eq. \eqref{eq: gendos}, 
resulting in a steeper energy spectrum as shown in Fig. \ref{fig: shoiso2mod}.

In brief, an oblique shock has larger $d$ and $P_\text{esp}$ than that of a parallel shock. 
For an oblique shock with a small shock speed,
the particle energy spectrum at small obliquities
has the same spectral index as that for a parallel shock, but 
becomes slightly shallower at large obliquities. 
When the shock speed is large, a very steep spectrum is expected for quasi-perpendicular shocks.

\section{Shock acceleration in the presence of dynamo-amplified magnetic fields}
\label{sec:turdyn}

 \subsection{Turbulent dynamo in the upstream and downstream regions}
 \label{ssec: turdyn}

The ubiquitous turbulence in the ISM induces density fluctuations over a broad range of length scales
\citep{Armstrong95,CheL10}.
In particular, the density spectra measured in the Galactic disk 
are much shallower than the Kolmogorov spectrum, indicative of 
an excess of small-scale density structures 
\citep{XuZ16,XuZ17,Xup20}. 
When an SN shock sweeps through density fluctuations 
\citep{Pais20},  
the interaction between the CR pressure gradient formed in the shock precursor
and upstream density inhomogeneities generates vorticity and thus turbulence in the upstream region, 
which further amplifies the interstellar magnetic fields via the turbulent dynamo 
\citep{BJL09,Dru12,Del16,XuL17}.
The turbulence is driven on the length scale of the typical size of density fluctuations $L$
and cascades toward smaller scales. 
{\xu The typical size of density fluctuations can be determined by the inner scale of a shallow density spectrum \citep{LP04}}.
The turbulent velocity $V_L$ at $L$ depends on both the relative density fluctuation $\delta \rho / \rho$
and the shock speed
\citep{Dru12}.

For initially super-Alfv\'{e}nic turbulence
with the turbulence Alfv\'{e}n Mach number $M_{A0} = V_L / V_{A0} >1$ in the shock precursor, 
where $V_{A0} = B_0 / \sqrt{4\pi\rho} $ is the initial Alfv\'{e}n speed corresponding to the 
initial interstellar magnetic field strength $B_0$,
the turbulent dynamo is in the nonlinear regime 
\citep{XL16}
(see \citealt{XuL17,Xupar19} for a different dynamo regime when an SN shock propagates
in a weakly ionized medium).
The magnetic energy $\mathcal{E}_M$ grows linearly with time
\citep{XL16}
\begin{equation}
     \mathcal{E}_M  = \mathcal{E}_{M0} + \frac{3}{38} \chi t, 
\end{equation}
where $\mathcal{E}_{M0} = 1/2 V_{A0}^2$ is the initial magnetic energy, 
and 
\begin{equation}
   \chi = L^{-1}V_L^3
\end{equation}
is the energy transfer rate along the turbulent energy cascade. 
Obviously, the upstream 
turbulence with a larger $V_L$ driven at a smaller $L$ can lead to a faster dynamo growth of magnetic energy, 
as tested in 
\citet{Del16}.
With the growth of $\mathcal{E}_M$, the length scale $l_A = 1/ k_A$
where $\mathcal{E}_M$ and the turbulent energy are in equipartition, 
also increases
\citep{XL16},
\begin{equation}
    k_A = \Big(k_{A0}^{-\frac{2}{3}}+ \frac{3}{19} \chi^\frac{1}{3} t\Big)^{-\frac{3}{2}} ,
\end{equation}
and $1/k_{A0}$ is the initial energy equipartition scale. 
$l_A$ is also the correlation length of turbulent magnetic fields and related to the turbulence Alfv\'{e}n Mach number
$M_A = V_L/ V_A$ by
\citep{Lazarian06}
\begin{equation}
   l_A = L M_A^{-3}, ~~~~M_A>1.
\end{equation}
If the nonlinear dynamo timescale 
\citep{XL16}
\begin{equation}\label{eq: nodytim}
     \tau_{dyn} \approx \frac{19}{3} \frac{L}{V_L}
\end{equation}
is shorter than the advection time through the precursor,
$\mathcal{E}_M$ can reach full equipartition with the turbulent energy $1/2 V_L^2$ at $L$, 
and $l_A = L$, i.e., $M_A =1$. 
Otherwise, the magnetic energy is 
$\mathcal{E}_M = 1/2 V_A^2$, with 
\begin{equation}\label{eq: dyvaca}
    V_A 
    = v_{l_A} = V_L \Big(\frac{l_A}{L}\Big)^\frac{1}{3},  ~~ l_A < L .
\end{equation}
Here $v_{l_A}$ is the turbulent velocity at $l_A$, and the Kolmogorov scaling of turbulent velocity is used 
{\sy within the inertial range of turbulence on scales larger than kinetic scales. The Kolmogorov spectrum of turbulence is seen in both simulations
(e.g., \citealt{Ino09})
and observations 
(e.g., \citealt{Shim18})
of SNRs. }

In the downstream region, if the magnetic energy is smaller than the turbulent energy, 
dynamo amplification of magnetic fields can also take place behind the shock front
\citep{Giac_Jok2007,Ino09,Ji16}. 
The time evolution of the postshock magnetic field and its strength indicated by 
the synchrotron X-ray emission of SNRs
\citep{Uch07,Ino09}
are found to be consistent with the expectation of the 
nonlinear turbulent dynamo theory 
\citep{XuL17}.

In super-Alfv\'{e}nic turbulence, due to the turbulent tangling of magnetic fields at $l_A$, 
CRs with $r_g \ll l_A$ 
have their effective mean free path as $l_A$
\citep{Brunetti_Laz}.
The corresponding energy-independent isotropic diffusion coefficient on scales larger than $l_A$ is
\begin{equation}\label{eq: brulaz}
   D = \frac{1}{3} l_A c.
\end{equation}
It applies when the scattering mean free path $\lambda_\|$ of CRs that interact with MHD turbulence 
is larger than $l_A$. 
Inefficient scattering by Alfv\'{e}nic turbulence was earlier studied by, e.g., 
\citet{Chan00,YL02,XL20}.
In compressible MHD turbulence, 
the mirroring effect can result in suppressed diffusion and 
significantly smaller mean free path than the scattering mean free path
\citep{LX21,Xu21}.   
When $\lambda_\|$ is smaller than $l_A$, there is 
\begin{equation}\label{eq: isescd}
    D = \frac{1}{3} \lambda_\| c,
\end{equation} 
which is also isotropic in super-Alfv\'{e}nic turbulence on scales larger than $l_A$
\citep{YL08}.
Here the energy dependence of $\lambda_\|$ is determined by the properties of compressible MHD turbulence 
\citep{LX21}.


\subsection{Spectral index of accelerated particles}

\begin{figure}[ht]
\centering
   \includegraphics[width=7.5cm]{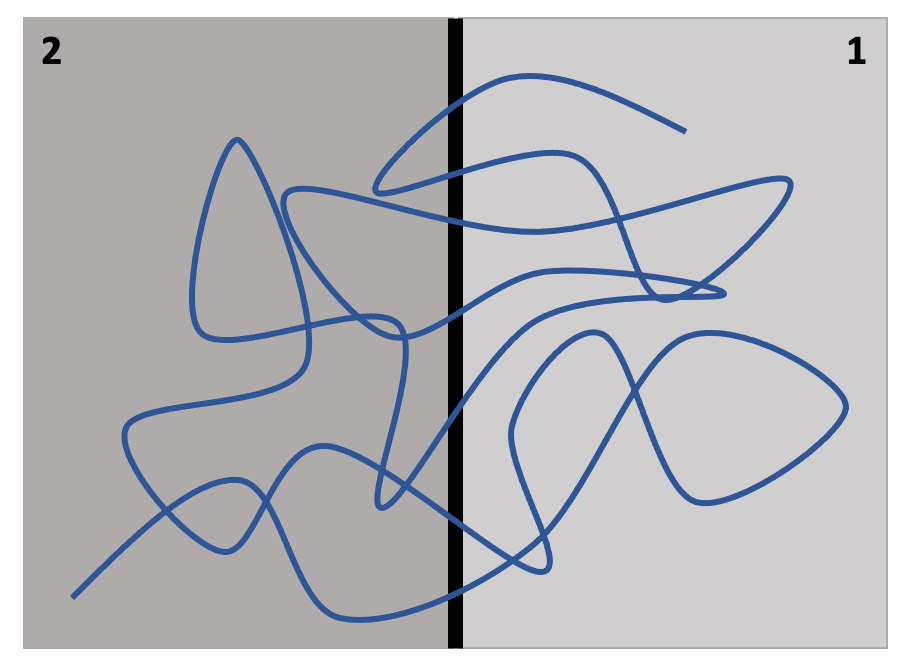}
\caption{Illustration of a shock in the presence of dynamo-amplified turbulent magnetic fields. ``1" and ``2" denote upstream and downstream regions.
{\sy The turbulent magnetic fields
(blue lines) result in 
a variety of shock obliquities
along the shock front (thick black line).} }
\label{fig: sket}
\end{figure}

For an SN shock propagating in the inhomogeneous ISM with density fluctuations, 
the dynamo-amplified turbulent magnetic fields in the shock’s vicinity result in 
a variety of shock obliquities, independent of the initial 
obliquity angle between the interstellar magnetic field and the shock normal (see Fig. \ref{fig: sket}). 
\footnote{We note that upstream density fluctuations can also cause a rippling of the shock surface 
and result in locally oblique shocks 
\citep{Giac_Jok2007}.}
The spatial extent of the local obliquity is given by the characteristic scale of turbulent magnetic fields
$l_A$. 
Here we consider the shock acceleration on scales larger than $l_A$.

In the case of efficient scattering with $r_g < \lambda_\| < l_A$, there is an 
isotropic pitch angle distribution at each locally oblique shock.
Based on the formulae for an oblique shock (see the expressions in Appendix \ref{sec:app}),
the mean fractional energy gain per cycle is 
\begin{equation}\label{eq: turdave}
    d(\alpha_1, \alpha_2^\prime) = P_{12}(\alpha_1) \Big[d_{12} (\alpha_1) + d_{21} (\alpha_2^\prime)\Big] + P_r (\alpha_1) d_r (\alpha_1).
\end{equation}
Here we note that because particles following turbulent field lines can interact with different oblique shock fronts at each encounter,
the obliquity angle $\alpha_i$ for $d_{r}$ and $d_{12}$ can be different from the obliquity angle $\alpha_i^\prime$ for $d_{21}$. 
After averaging over all possible obliquities for subluminal shock fronts, we further obtain
\begin{equation}\label{eq: turavobld}
   d = \frac{\iint d(\alpha_1,\alpha_2^\prime) d\cos\alpha_1 d \cos\alpha_2^\prime }{\int d \cos \alpha_1 \int d\cos\alpha_2^\prime} .
\end{equation}
The obliquity averaged probability of escape is 
\begin{equation}
   P_\text{esp} = \overline{P_{12}} \frac{\overline{F_2^\text{esp}}}{\overline{F_{2}}}\approx \overline{P_{12}} \frac{8 U_2}{c} =  \overline{P_{12}} \frac{2 U_1}{c},
\end{equation} 
where $\overline{...}$ denotes the average over obliquities, 
$F_2^\text{esp}$ is the flux of escaping particles, 
$F_2$ is the flux of particles that enter into the downstream region, 
and (Eq \eqref{eq: fnetfgen})
\begin{equation}
      \overline{F_2^\text{esp}} \approx n_2 U_2,  ~~\overline{F_{2}} \approx \frac{n_2 c}{8}.
\end{equation}
$n_2$ is the particle number density in the downstream fluid frame.

\begin{figure*}[ht]
\centering
\subfigure[]{
   \includegraphics[width=8.7cm]{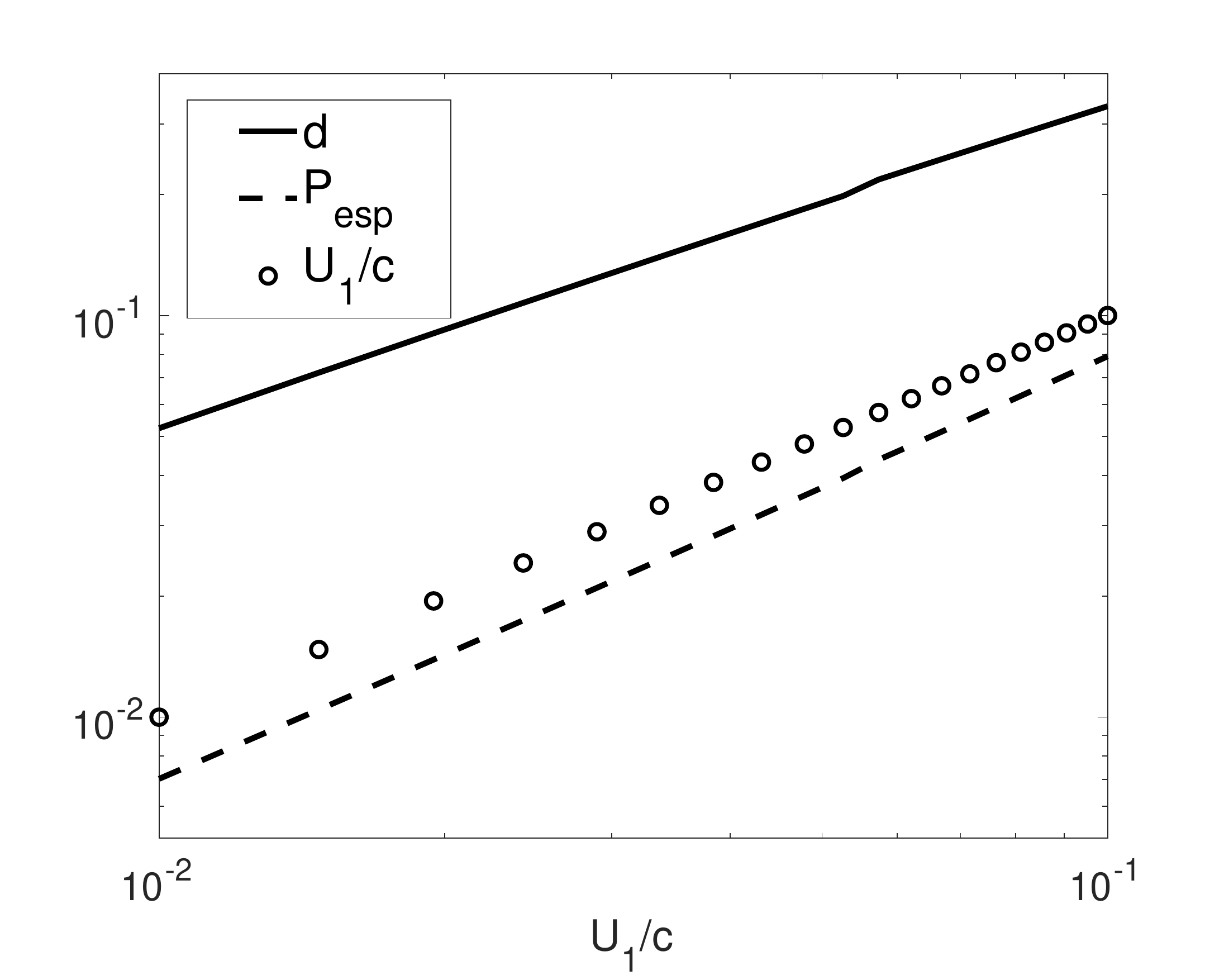}\label{fig: turshodp}}
\subfigure[]{
   \includegraphics[width=8.7cm]{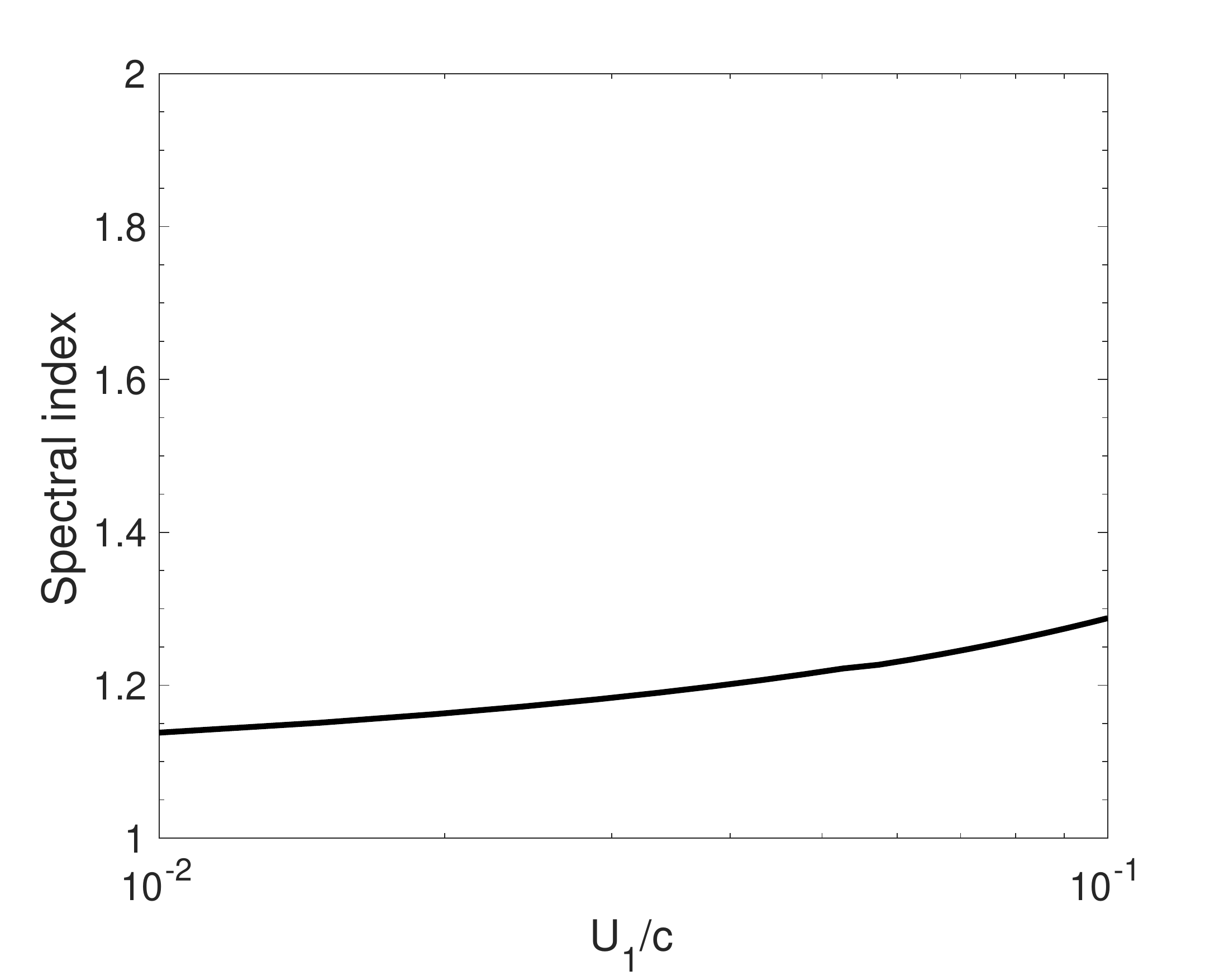}\label{fig: turshospe}}
\caption{Obliquity-averaged $d$ and $P_\text{esp}$ in (a) and the corresponding spectral index in (b) for 
shock acceleration at scales larger than $l_A$ 
in the presence of dynamo-amplified turbulent 
magnetic fields. 
Efficient scattering and isotropic pitch angle distribution are assumed.  
$U_1/c$ (circles) is shown in (a) as a comparison.
{\sy We note that the little bump seen in the profile of $d$ is an artifact from the numerical evaluation of Eq. \eqref{eq: turavobld}.}}
\label{fig: turshoc}
\end{figure*}

In Fig. \ref{fig: turshoc}, we present the obliquity averaged $d$, $P_\text{esp}$, and the spectral index of accelerated particles. 
We see that both $d$ and $P_\text{esp}$ increase with the shock speed.
$d$ is larger than $U_1/c$, which is both mean fractional energy gain and escape probability for a parallel shock
(see Section \ref{sec: paralsh}), but smaller than that for a highly oblique shock
(Figs. \ref{fig: dpesp} and \ref{fig: dpesp2mod}).
The small $P_\text{esp}$
justifies the assumption of negligible loss of particles when we calculate $d(\alpha_1, \alpha_2^\prime)$ in Eq. \eqref{eq: turdave}. 
With $d > P_\text{esp}$, 
a very shallow spectrum is found, and the spectral index has a weak dependence on the shock speed.

\begin{figure*}[ht]
\centering
\subfigure[]{
   \includegraphics[width=8.7cm]{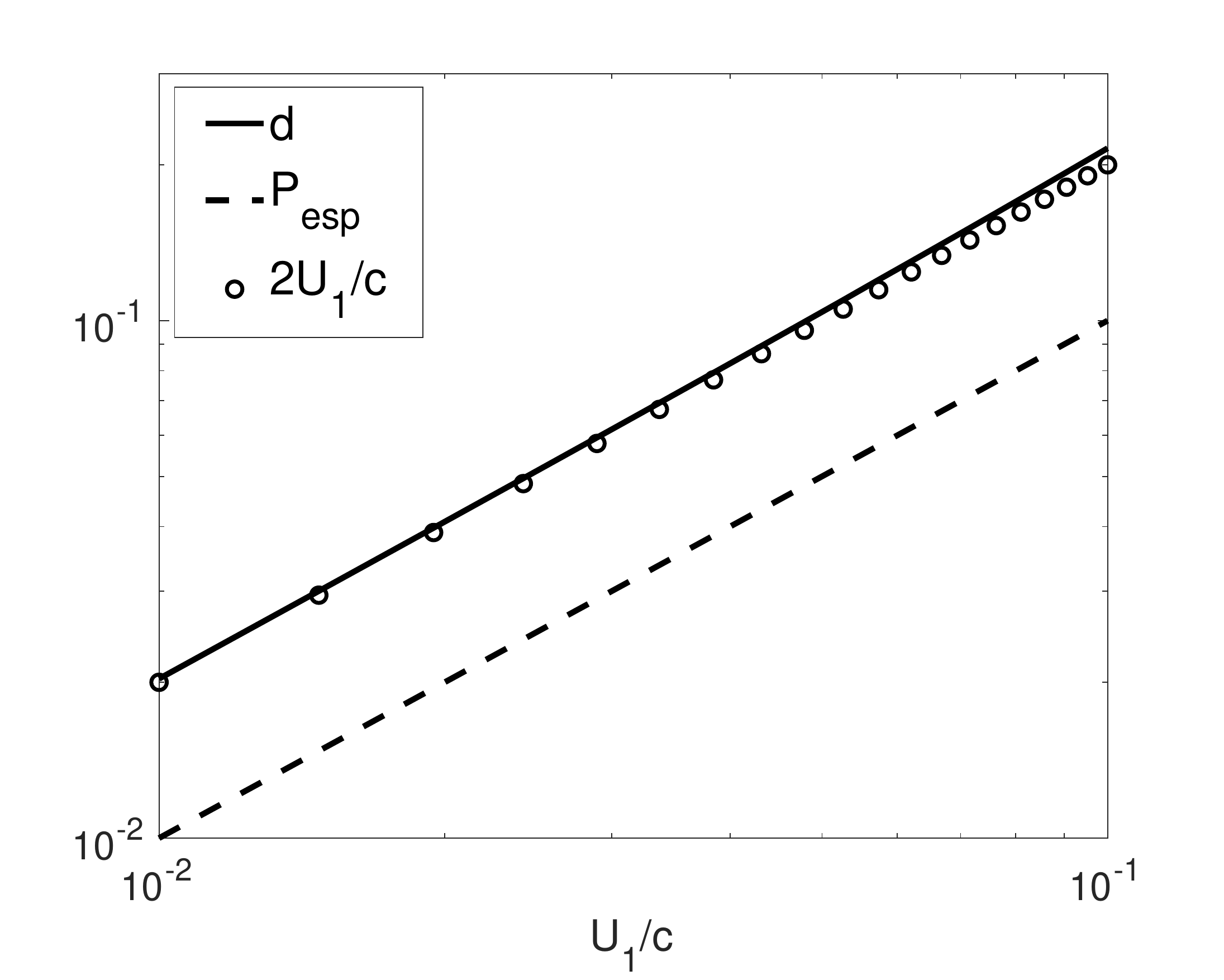}\label{fig: turshom1d}}
\subfigure[]{
   \includegraphics[width=8.7cm]{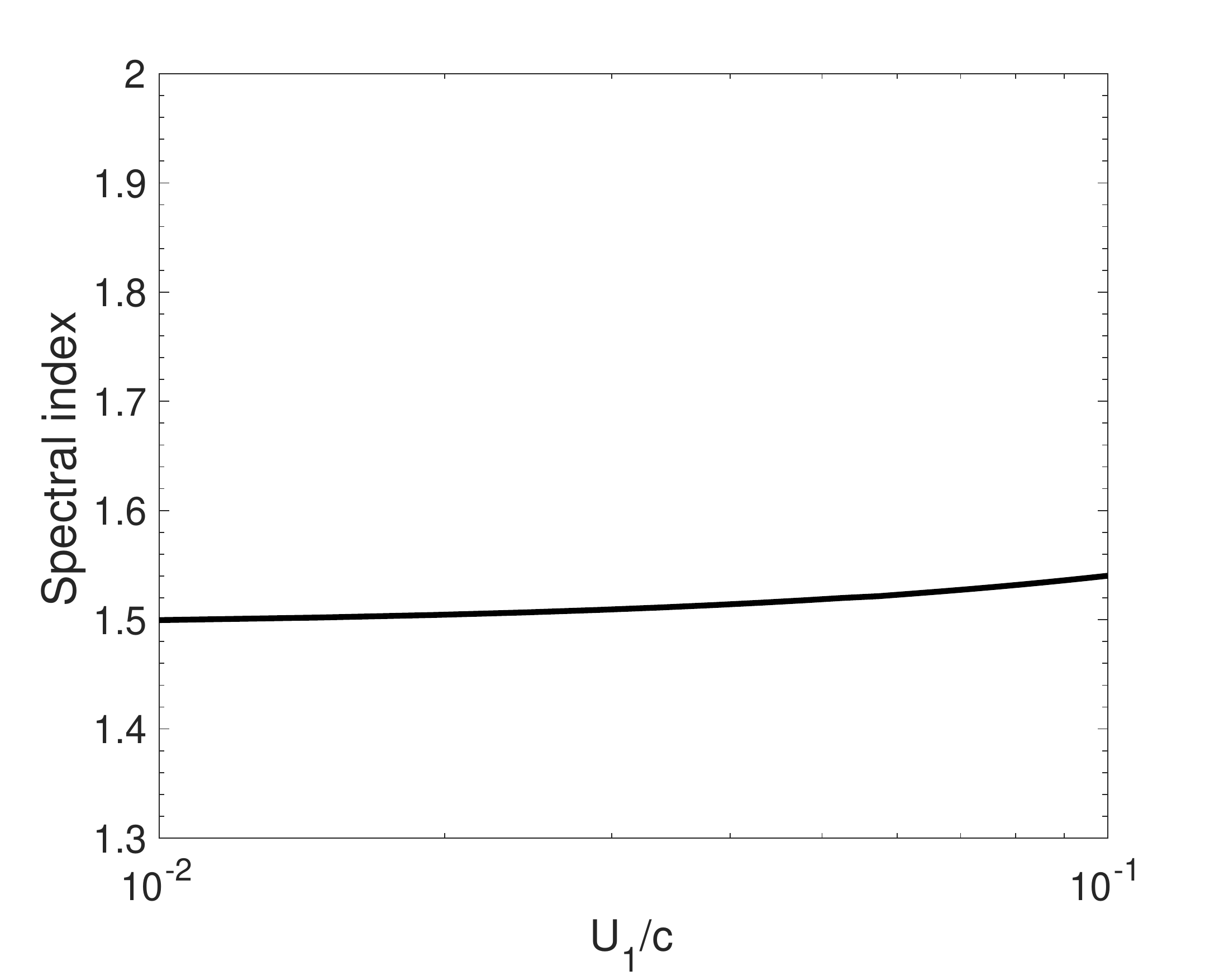}\label{fig: turshospem1}}
\caption{ Same as Fig. \ref{fig: turshoc} but for inefficient scattering and $\mu=1$.
Circles in (a) represent the approximate $d$ given by Eq. \eqref{eq: appdtum1}.}
\label{fig: turshocm1}
\end{figure*}

On the one hand, turbulent magnetic fields enable larger energy gain than a parallel shock by creating 
locally oblique shock fronts.  
On the other hand, they lead to isotropization of the spatial distribution of particles irrespective of scattering and thus 
decrease the probability of particle loss 
in the downstream region. 
As a result, a shallow spectrum is expected. 
The effect of dynamo-amplified magnetic fields on shock acceleration can be more easily seen in the case of  
inefficient scattering with the scattering mean free path larger than $l_A$. 
In this case, particles following turbulent magnetic fields have an effective mean free path equal to $l_A$ (Section \ref{ssec: turdyn}).
As an example, we consider $\mu = 1$ for all particles that ballistically move along field lines,
where $\mu$ is cosine of particle pitch angle.  
Without reflection, particles undergo two transmissions per cycle. 
The mean fractional energy gain during the transmission is 
\begin{equation}
\begin{aligned}
   &d_{12} = \frac{\int V_{rel,1} \Big[\Big(\frac{E_f}{E_0}\Big)_{12}-1\Big] d \cos\alpha_1}{ \int V_{rel,1} d \cos \alpha_1},\\
   &d_{21} = \frac{\int V_{rel,2} \Big[\Big(\frac{E_f}{E_0}\Big)_{21}-1\Big] d \cos\alpha_2^\prime}{ \int V_{rel,2} d \cos\alpha_2^\prime}.
\end{aligned}
\end{equation}
Then we have 
\begin{equation}\label{eq: m1tud}
    d = d_{12} + d_{21} .
\end{equation} 
Despite the inefficient scattering, particles are coupled to the fluid by moving along turbulent magnetic fields and 
have isotropic spatial distribution on scales larger than $l_A$. 
As all particles are transmitted through the shock,
we can approximately treat it as a parallel shock, with $U_1$ adjusted due to the presence of turbulent magnetic fields, 
and have 
\begin{equation}\label{eq: appdtum1}
    d \approx \frac{\overline{U_1}}{c} = \frac{U_1}{c \int \cos \alpha_1 d\cos\alpha_1} \approx \frac{2 U_1}{c}.
\end{equation}
Similar to the case of a parallel shock, there is also
\begin{equation}
    P_\text{esp} = \frac{U_1}{c}.
\end{equation}
The corresponding results are shown in Fig. \ref{fig: turshocm1}.
The expression in Eq. \eqref{eq: appdtum1} provides a good approximation of $d$.
$d$ becomes smaller than that for an isotropic pitch angle distribution 
due to the absence of reflection (see Fig. \ref{fig: turshodp}). 
The resulting spectrum index is still shallower than that for a parallel shock 
because 
$d>P_\text{esp}$.

\section{Acceleration time}
\label{sec: acctime}
 
\subsection{Acceleration time for parallel and oblique shocks}

 
 
 
The acceleration time is defined as 
\citep{Drury83}
\begin{equation}\label{eq: satcc}
    t_\text{acc} = \frac{\Delta t}{d}. 
\end{equation} 
For an oblique shock, the mean cycle time of a particle is 
\citep{Ostr88}
\begin{equation}\label{eq: oshoct}
   \Delta t =  P_r  t_1 + P_{12} (t_1 + t_2)  ,
\end{equation}  
where 
\begin{equation}
   t_1 = \frac{2 D_{n1}}{v_{n1} U_1},   ~~ t_2 = \frac{2 D_{n2}}{v_{n2} U_2}
\end{equation}
are the mean residence time in the upstream and downstream regions,
and $D_{ni}$ and $v_{ni}$ are the diffusion coefficient and the mean particle velocity along the shock normal. 
$D_{ni}$ can be expressed in terms of the parallel ($D_{\| i}$) and perpendicular ($D_{\perp i}$) diffusion coefficients
\citep{Jok87}
\begin{equation}\label{eq: dffjok}
    D_{ni} = D_{\| i} \cos^2 \alpha_i + D_{\perp i} \sin^2 \alpha_i.
\end{equation}
The expression of $v_{ni}$ is given by 
\citep{Ostr88}
\begin{equation}\label{eq: vneos}
   v_{ni} = v_\|\sqrt{\frac{D_{ni}}{D_{\| i}}},
\end{equation}
where $v_\|$ is the mean particle velocity along the magnetic field.


(1) Parallel shock. 

For a parallel shock, no magnetic filed compression takes place, so we have 
\begin{equation}
    P_r = 0, ~~ P_{12} = 1,
\end{equation}
and (Eqs. \eqref{eq: pasbe}, \eqref{eq: dffjok}, and \eqref{eq: vneos})
\begin{equation}
    d = \frac{U_1}{c}, ~~ D_{ni} = D_{\| i}, ~~ v_{ni} = v_\|. 
\end{equation}
Then Eqs. \eqref{eq: satcc} and \eqref{eq: oshoct} give 
\begin{equation}\label{eq: taccpar}
\begin{aligned}
   t_\text{acc} &= \frac{t_1+t_2}{d}\\
               &= \frac{2c}{v_\| U_1} \Big(  \frac{ D_{\|1}}{ U_1} +  \frac{ D_{\|2}}{ U_2} \Big) \\
               & =  \frac{10 D_\| c}{v_\| U_1^2}  \\
               & = \frac{20 D_\| }{ U_1^2}  ,
\end{aligned}
\end{equation}
where $D_{\| 1} = D_{\| 2} = D_\|$ is assumed, and $v_\| = 1/2 c$ for isotropic particle distribution. 
With a small energy gain at each shock crossing, 
efficient acceleration for a parallel shock is expected only when the scattering is efficient and $D_\|$
is sufficiently small.

(2) Oblique shock with highly anisotropic diffusion 

For highly anisotropic diffusion with $D_{\| i} \gg D_{\perp i}$, there is (Eqs. \eqref{eq: dffjok} and \eqref{eq: vneos})
\begin{equation}\label{eq: slnove}
    v_{ni} \approx v_\| \cos \alpha_i 
\end{equation}
for most obliquities. 
So the acceleration time is 
\begin{equation}\label{eq: obandiff}
\begin{aligned}
   t_\text{acc} & = \frac{1}{d} (t_1 + P_{12} t_2) \\
                      & \approx \frac{2}{d v_\|}  \Big(\frac{ D_{n1}}{ \cos \alpha_1 U_1} + P_{12} \frac{ D_{n2}}{ \cos \alpha_2 U_2} \Big) 
\end{aligned}
\end{equation}
except for quasi-perpendicular shocks. 
To the approximation order $O(\beta_1)$ 
with $d \approx \beta_1$ and $P_{12} \approx 1-\mu_0^2$,
the above expression can be simplified as 
\begin{equation}
\begin{aligned}
   t_\text{acc}  
                       & \approx \frac{2c}{ v_\|  U_1^2}  ( D_{n1} +   4D_{n2} )  \\
                       & = \frac{4}{   U_1^2}  ( D_{n1} +   4D_{n2} ) ,
\end{aligned}
\end{equation}
where we use the relation in Eq. \eqref{eq: mu0bb}. 
This is the same as the acceleration time derived in 
\citet{Jok87}. 
Although the energy gain per cycle 
increases with obliquity, 
$v_{ni}$ (Eq. \eqref{eq: slnove}) decreases with increasing obliquity 
as the diffusion of particles is mainly along the magnetic field.  
As a result, the obliquity does not directly enter the approximate expression of $t_\text{acc}$.
\citet{Jok87}
argued that for a quasi-perpendicular shock, given $D_{\perp i} \ll D_{\| i}$, 
the rate of acceleration, i.e., $1/t_\text{cas}$, can be much faster than that of a quasi-parallel shock.


(3) Oblique shock with sub-Alfv\'{e}nic turbulence

In \citet{Jok87}, 
the standard kinetic relationship between perpendicular and parallel diffusion was applied, i.e., 
$D_{\perp i } = D_{\| i}/ (1 + (\lambda_\|/ r_g)^2)$,
leading to highly suppressed perpendicular diffusion when $\lambda_\| \gg r_g$.
In turbulent magnetic fields, the perpendicular (super)diffusion of particles is regulated by the (super)diffusion of turbulent magnetic fields. 
The particle perpendicular (super)diffusion in different turbulence regimes has been theoretically studied and numerically tested with 
test particle simulations 
\citep{XY13,LY14,LY19}.

\begin{figure}[ht]
\centering
   \includegraphics[width=7cm]{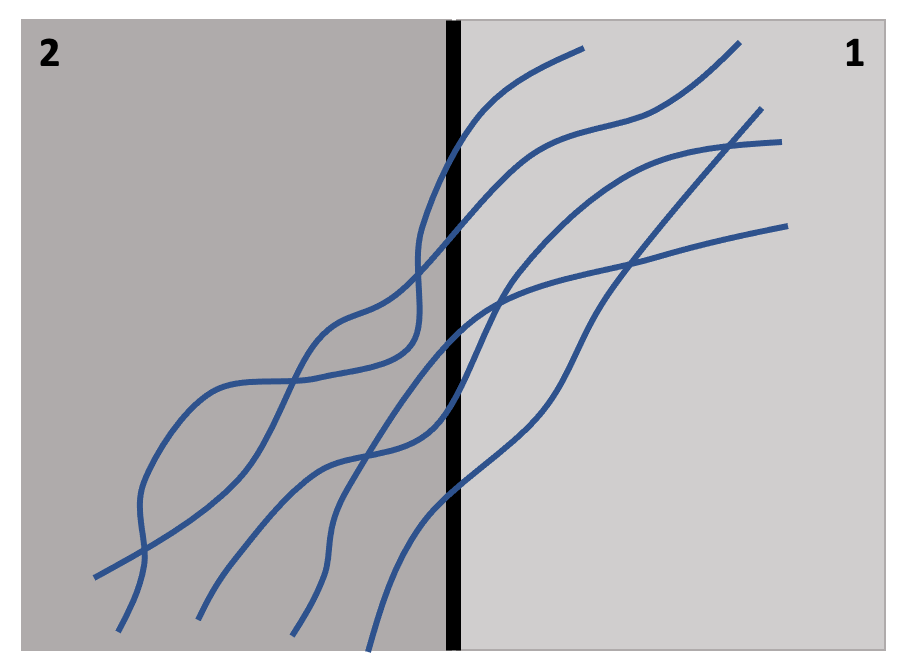}
\caption{Sketch for an oblique shock with mildly turbulent magnetic fields {\sy (blue lines)} in sub-Alfv\'{e}nic turbulence.}
\label{fig: turobsek}
\end{figure}

For an oblique shock with sub-Alfv\'{e}nic turbulence, the turbulence amplitude is relatively small so that 
local obliquity angles are not very different from the global one between the shock normal and the upstream mean magnetic field
(see Fig. \ref{fig: turobsek}).
If the injection scale $L$ of turbulence is sufficiently large, perpendicular superdiffusion 
of turbulent magnetic fields is expected on scales smaller than $l_\text{tr} = LM_A^2$ ($M_A<1$), 
which is the transition scale from the weak turbulence over the range $[l_\text{tr},L]$ to the strong turbulence on smaller length scales. 
The effect of perpendicular superdiffusion of magnetic fields and particles on shock acceleration was earlier discussed in 
\citet{LY14}.
If the upstream turbulence is driven at a small $L$, the perpendicular diffusion of particles 
on scales larger than $L$ is related to the parallel diffusion by 
\citep{YL08,XY13}
\begin{equation}
    D_\perp = D_\| M_A^4, ~~ M_A<1, ~~ \lambda_\| < L.
\end{equation}
Note that this scaling is different from the scaling $D_\perp \propto M_A^2$ resulting from the field line wandering 
based on the quasi-linear theory 
(QLT; \citealt{Jokipii1966,JokPar68}).
At a small $M_A$, we also have highly anisotropic diffusion in sub-Alfv\'{e}nic turbulence, and thus Eq. \eqref{eq: obandiff} 
for $t_\text{acc}$ applies except for quasi-perpendicular shocks.

\subsection{Acceleration time for a shock in the presence of dynamo-amplified turbulent magnetic fields}

In super-Alfv\'{e}nic turbulence with dynamo-amplified magnetic fields,
the diffusion is isotropic (see Section \ref{ssec: turdyn}).
With $D_{\| i} = D_{\perp i} = D_i$, there is $D_{ni} = D_i$ (Eq. \eqref{eq: dffjok}), and 
$v_{ni} = v_\|$ (Eq. \eqref{eq: vneos}).
$t_\text{acc}$ can be expressed as 
\begin{equation}\label{eq: tacdystsup}
\begin{aligned}
   t_\text{acc} & = \frac{1}{d} (t_1 + \overline{P_{12}} t_2) \\
                      & = \frac{2}{d v_\|} \Big(\frac{ D_1}{ U_1} + \overline{P_{12}} \frac{D_2}{ U_2}\Big) \\
                      & = \frac{4 D}{d c U_1} \big(1 +  4 \overline{P_{12}} \big) \\
                      & < \frac{20 D}{d c U_1} , 
\end{aligned}
\end{equation}
where we assume $D_1 = D_2 = D = 1/3 \lambda_\| c$ (Eq. \eqref{eq: isescd}) and isotropic pitch-angle distribution, and 
$d$ is given by Eq. \eqref{eq: turavobld}.

If $\mu = 1$, then there is 
\begin{equation}
   t_\text{acc}  = \frac{1}{d} (t_1 +  t_2) 
    = \frac{10 D}{dc U_1} ,
\end{equation}
where $D$ and $d$ are given by Eq. \eqref{eq: brulaz} and Eq. \eqref{eq: m1tud}, respectively. 
As $d > U_1/c$ (Figs. \ref{fig: turshodp} and \ref{fig: turshom1d}), $t_\text{acc}$ for a shock with dynamo-amplified turbulent magnetic fields 
is in general smaller than that of a parallel shock.

\begin{figure}[ht]
\centering
   \includegraphics[width=8.5cm]{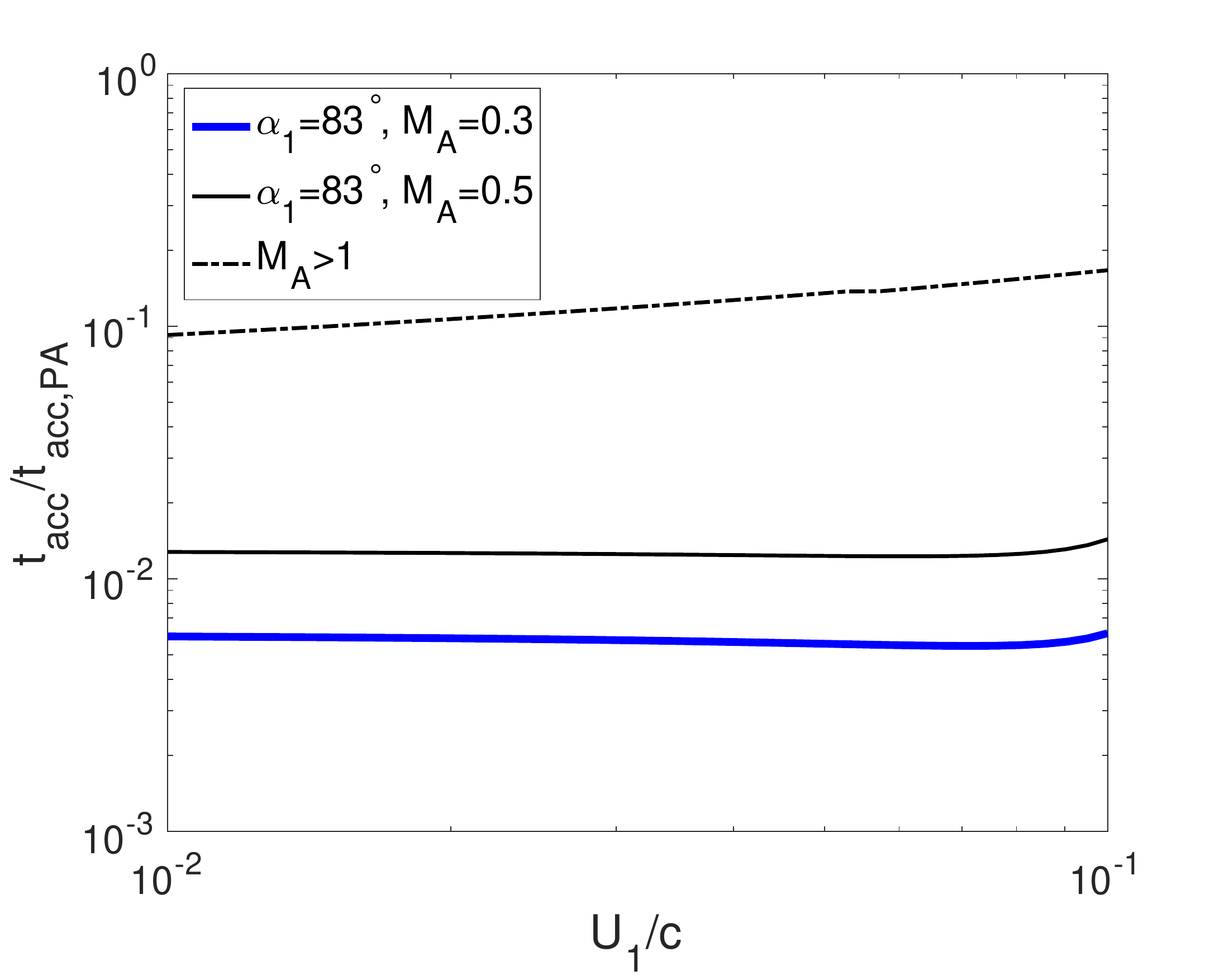}
\caption{$t_\text{acc}$ of a quasi-perpendicular shock with sub-Alfv\'{e}nic turbulence at $M_A = 0.3$ (blue thick solid line)
and at $M_A = 0.5$ (thin solid line)
normalized by $t_\text{acc,PA}$ of a parallel shock, 
and $t_\text{acc}$ of a shock with dynamo-amplified magnetic fields in super-Alfv\'{e}nic turbulence (dash-dotted line) 
normalized by $t_\text{acc,PA}$.}
\label{fig: acctime}
\end{figure}

In Fig. \ref{fig: acctime}, we compare $t_\text{acc}$ for isotropic pitch-angle distribution
of a parallel shock, i.e., $t_\text{acc,PA}$ (Eq. \eqref{eq: taccpar}), 
a quasi-perpendicular shock with sub-Alfv\'{e}nic turbulence (Eq \eqref{eq: gendoblo}, Eqs \eqref{eq: satcc}-\eqref{eq: vneos}),
and a shock with dynamo-amplified magnetic fields in super-Alfv\'{e}nic turbulence (Eqs. \eqref{eq: turavobld} and \eqref{eq: tacdystsup}). 
Here for simplicity, 
we assume $D_{\|i} = D_\|$ for a parallel shock, 
$\alpha_1 = 83^\circ$, $D_{\perp i}=D_{\| i} M_{Ai}^4 = D_{\|} M_A^4$ for a quasi-perpendicular shock, 
and $D_i = D_\|$ for a shock with super-Alfv\'{e}nic turbulence, 
but note that the diffusion coefficients and $M_A$ vary with turbulence properties and 
can be different in upstream and downstream regions in reality. 
Compared to the DSA for a parallel shock, 
we see more efficient acceleration for a quasi-perpendicular shock with anisotropic diffusion and 
a shock with dynamo-amplified turbulent magnetic fields and isotropic diffusion, 
and $t_\text{acc}$ depends on $M_A$ of the upstream turbulence. 
  

\section{Discussion}

\subsection{Acceleration efficiency of quasi-perpendicular and quasi-parallel shocks} 

We find a more efficient acceleration at a quasi-perpendicular shock than at a quasi-parallel shock under the condition that magnetic fields are turbulent in both scenarios, and the turbulence enables the diffusion of particles. 
For an SN shock propagating through tenuous ambient medium, e.g., SN1006,
which is $500$ pc above the Galactic plane
\citep{Acer07},
dynamo amplification can be inefficient given the small density inhomogeneities.
In this case, a quasi-parallel shock is more favored for 
the generation of magnetic fluctuations via CR-induced instabilities
and particle acceleration
\citep{Caps14}, 
as suggested by polarization and synchrotron measurements of SN 1006
\citep{Kats17}.
For SNRs in an inhomogeneous environment in the Galactic disk,
the dynamo-amplified turbulent magnetic fields contribute to both parallel and perpendicular (super)diffusion of CRs, 
and thus quasi-perpendicular shocks with anisotropic diffusion can lead to more efficient acceleration. 
The acceleration efficiency of a shock depends on not only the shock obliquity but also the interstellar environment. 
New techniques have been developed for measuring the turbulent magnetic fields in the vicinity of SNRs 
(e.g., \citealt{Lium21}).

{\sy For interplanetary shocks in the solar wind, the shocks at 1 AU are mainly quasi-perpendicular with respect to the Parker spiral magnetic field
\citep{Guo21}. 
The ambient solar-wind turbulence plays an important role in facilitating diffusion of particles.
As a result, 
quasi-perpendicular shocks 
can be more efficient in accelerating particles and 
are more geoeffective than quasi-parallel
shocks
(e.g., \citealt{Jur02,Guo21}).
Different from 
the interstellar turbulence, 
the solar wind turbulence can give rise to a significant level of upstream magnetic fluctuations  
at scales smaller than the 
spatial extent of interplanetary shocks
\citep{Pit21}, 
which can affect the  
diffusion of particles even when the  
upstream turbulent dynamo is inefficient. }

\subsection{Dynamo amplification of magnetic fields}

\begin{figure}[ht]
\centering
   \includegraphics[width=8cm]{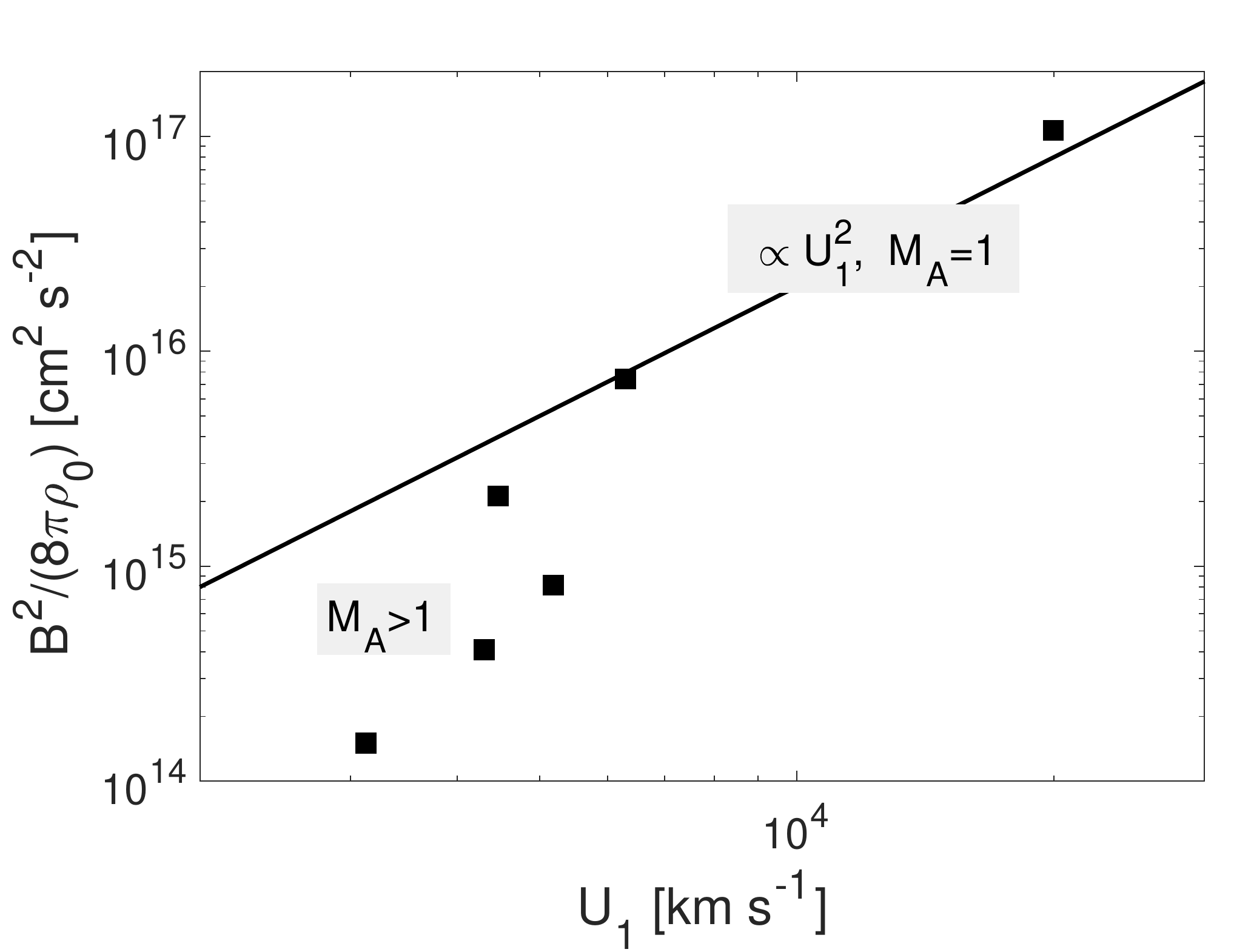}
\caption{Post-shock magnetic energy density vs. shock speed. The data points are taken from \citet{Vin08}, where $\rho_0$ is the ISM density, and 
$B$ is the post-shock magnetic field strength. 
The solid line shows a dependence $\propto U_1^2$, corresponding to the saturated magnetic energy of the nonlinear turbulent dynamo.}
\label{fig: megush}
\end{figure}

For dynamo-amplified turbulent magnetic fields, the maximum magnetic energy at the the saturation of nonlinear turbulent dynamo is 
\begin{equation}
    \mathcal{E}_{M,\text{sat}} = \frac{1}{2} V_L^2. 
\end{equation}
With the equipartition between the magnetic energy and turbulent energy at $L$, the corresponding turbulence $M_A$ is unity. 
By adopting a crude approximation 
$V_L \propto U_1$, we get $\mathcal{E}_{M,\text{sat}}\propto U_1^2$.
In Fig. \ref{fig: megush},
we compare the above scaling with the measurements by 
\citet{Vin08},
where the magnetic fields are determined from X-ray synchrotron emission of young SNRs. 
We see that toward a slower shock speed, the magnetic energy is lower than the saturated value with $M_A>1$.
In this case, the advection time through the precursor is not sufficient for the nonlinear dynamo to reach full energy equipartition at $L$
(see Section \ref{ssec: turdyn}).
A different scaling with $B^2 \propto \rho_0 U_1^3$ was proposed to explain the data based on the CR-driven non-resonant 
growing modes at wavelengths shorter than the CR gyroradius
\citep{Bell2004}.
Compared with the CR current-driven instability,
the dynamo amplification of magnetic fields takes place in the upstream region with inhomogeneous density 
distribution.
The characteristic scale of dynamo-amplified magnetic fields is determined by the 
characteristic scale of density fluctuations and thus can be much larger than 
the CR gyroradius. 
Both processes can operate to amplify magnetic fields in SNRs. 
Their relative importance in different physical conditions deserves a more detailed study.

\subsection{Steep and flat radio spectra of SNRs}

\begin{figure}[ht]
\centering
   \includegraphics[width=9cm]{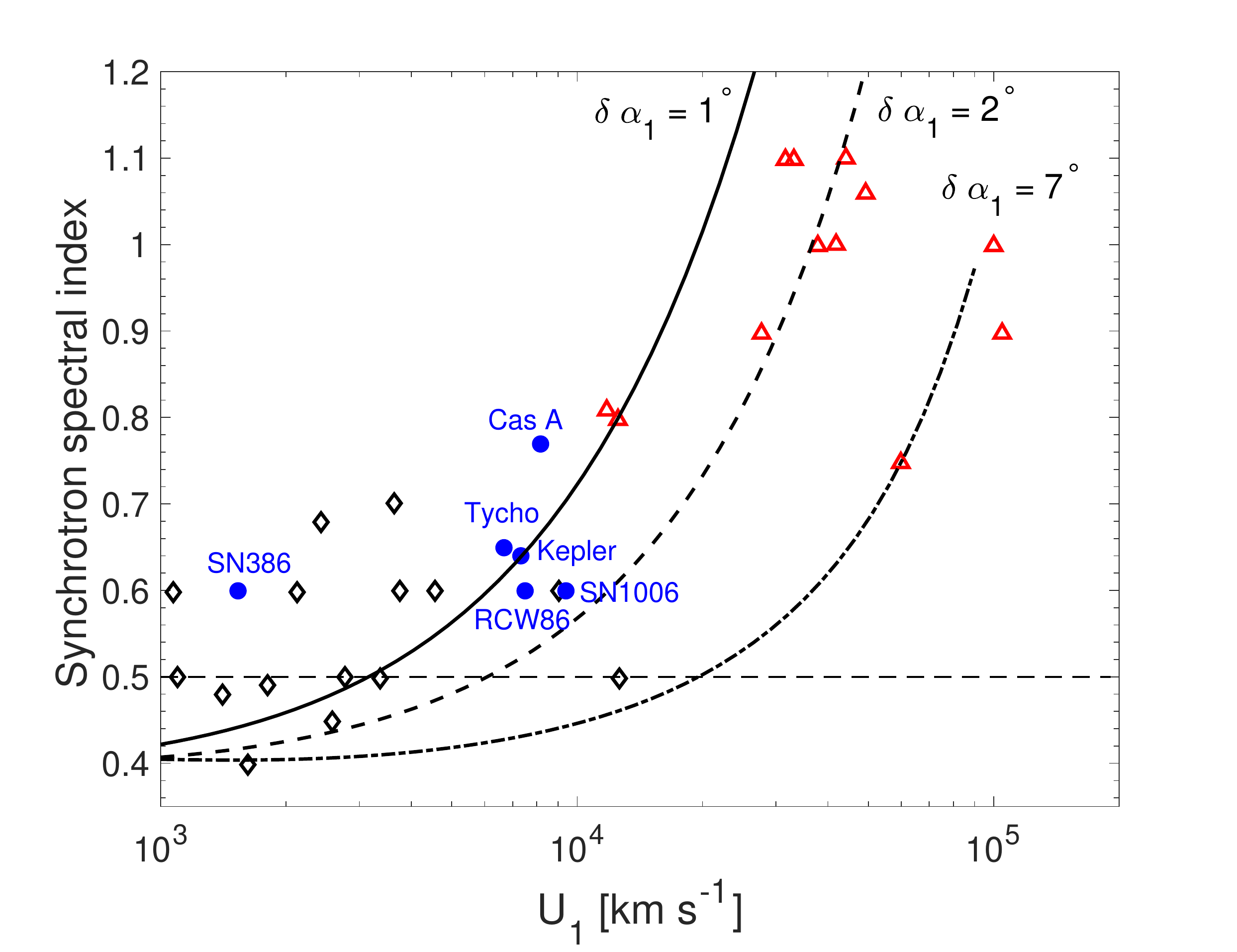}
\caption{Radio synchrotron spectral index vs. shock speed for young SNRs, including 
Galactic (diamonds), historic (circles), and extragalactic (triangles) sources.
The data points are taken from 
\citet{Be21}.
The horizontal dashed line corresponds to the index expected for a parallel shock. 
The three curves are from our calculations for the obliquity-averaged synchrotron spectral index of 
a quasi-perpendicular shock with a range of obliquities $[\alpha_{1,\text{max}}-\delta \alpha_1, \alpha_{1,\text{max}}]$, 
where $\alpha_{1,\text{max}}$ is the maximal $\alpha_1$ for a subluminal shock, and $\delta \alpha_1$ is the obliquity variation
caused by magnetic fluctuations. }
\label{fig: comobs}
\end{figure}

Observations show steep radio spectra of young SNRs, with the spectral indices of 
accelerated particles larger than $2$
\citep{Uro14}.
As shown in Fig. \ref{fig: comobs}, where the data points are taken from 
\citet{Be21},
young SNRs with large expansion velocities tend to have steep radio synchrotron spectra. 
The synchrotron spectral index (SI) is $=(\text{particle energy spectral index} -1) /2$. 
So $ \text{SI}=0.5$ corresponds to the case of a parallel shock with DSA (Eq. \eqref{eq: pwspec}), 
which cannot explain the spectral steepening at large shock speeds. 
Based on the results in Section \ref{sec: acctime}, 
a shock with a large obliquity 
is most efficient in accelerating particles (see Fig. \ref{fig: acctime}). 
If we assume that the acceleration is dominated by quasi-perpendicular shocks with the largest possible obliquities for subluminal shocks, 
we obtain the SI that increases with increasing $U_1$. 
By adjusting the obliquity variation $\delta \alpha_1$ that can be caused by magnetic fluctuations, 
we calculate the obliquity-averaged spectral index and 
find that with a larger $\delta \alpha_1$, significant spectral steepening is seen at a larger $U_1$.
Given that the shock speed can be underestimated on average 
\citep{Be21}, 
a larger $\delta \alpha_1$ is expected for some young SNRs.

Steep spectra of quasi-perpendicular shocks at large shock speeds were earlier suggested by, e.g, 
\citet{NaTa95,Be21}.
Here we stress that a quasi-perpendicular shock can be found in both cases with a uniform magnetic field 
and dynamo-amplified turbulent magnetic fields. 
Low-energy radio electrons can sample 
a local quasi-perpendicular shock in the presence of turbulent magnetic fields. 
As magnetic fluctuations decrease with decreasing length scales due to the turbulence cascade, the small local magnetic fluctuation $\delta b$ gives  
$\tan (\delta \alpha_1) \sim \delta b / B_0$, where $B_0$ is the mean magnetic field strength.

Old SNRs expanding in a high density environment and interacting with molecular clouds (MCs)
are observed to have flat radio spectra, with the spectral indices of 
accelerated particles smaller than $2$
\citep{Oni13}.
Due to the upstream density inhomogeneity in a high-density medium, vorticity-driven turbulence 
and dynamo-amplified magnetic fields are expected in both upstream and downstream regions, leading to 
shallow energy spectra of accelerated particles
(see Section \ref{sec:turdyn}). 
Interestingly, it is found that there is a significant connection between the SNRs with flat radio spectra and their detection in gamma-rays
\citep{Oni13,Ace16}.
The MCs in the vicinity of these SNRs serve as the target material for the shock-accelerated protons, 
resulting in gamma-ray emission via neutral pion decay. 
The steep gamma-ray spectra commonly seen from the SNR-MC systems can be explained by the mirror diffusion 
of CRs in compressible MHD turbulence in the MCs 
\citep{Xu21}.

\subsection{Maximum energy of accelerated particles}

The maximum energy of accelerated particles can be limited by the energy loss, the finite age and finite size of the source 
\citep{Lag83,Ach00}.
Here we consider the maximum energy limited by the 
characteristic scale of dynamo-amplified turbulent magnetic fields. 
The accelerated particles cannot be further confined by turbulent magnetic fields when $r_g$ becomes comparable to the 
characteristic scale $l_A$ of magnetic fields. 
By equalizing $r_g$ and $l_A$, 
\begin{equation}\label{eq: conmaxene}
   r_g 
         \approx \frac{E_\text{max}}{q B}
         = l_A, 
\end{equation}
we can obtain the corresponding maximum energy of accelerated protons, 
\begin{equation}
\begin{aligned}
   E_\text{max}& = e l_A  B  = e \sqrt{4\pi\rho} V_L l_A  \Big(\frac{l_A}{L}\Big)^\frac{1}{3} \\
                       &  \approx 10^{15} \text{eV}  \Big(\frac{n_H}{0.1 ~\text{cm}^{-3}}\Big)^\frac{1}{2}
                       \Big(\frac{V_L}{10^5 ~ \text{m s}^{-1}}\Big) \\
                    & ~~~~   \Big(\frac{l_A}{0.1 ~\text{pc}}\Big)^\frac{4}{3} \Big(\frac{L}{0.1 ~ \text{pc}}\Big)^{-\frac{1}{3}},
\end{aligned}
\end{equation}
where Eq. \eqref{eq: dyvaca} is used, and $n_H$ is the number density of atomic hydrogen. 
Obviously, a smaller $E_\text{max}$ is expected for a smaller $l_A$.
For SN shocks with dynamo-amplified turbulent magnetic fields, 
the condition in Eq. \eqref{eq: conmaxene} should be compared with other conditions that limit the particle energy, 
and $E_\text{max}$ is determined by the most stringent condition.



\subsection{Scattering of particles}

In most cases studied in this work, 
we assume efficient pitch angle scattering and isotropic pitch angle distribution. 
Efficient scattering can result from the interaction of particles with compressible MHD turbulence
(e.g., \citealt{XY13,XLb18,XL20})
and CR-driven instabilities 
(e.g., \citealt{Skil75,Bell2004,Cap14,vanM18,Bai19}).
If the scattering is so efficient that $\lambda_\|$ is comparable to $r_g$, 
the particle trajectory is significantly perturbed and 
the invariance of magnetic moment at the shock encounter 
is violated. 
\citet{Deck88}
showed that the introduction of magnetic fluctuations in general increases the energy gain of particles during the 
SDA.

When the scattering is inefficient,  
particles following turbulent magnetic fields in super-Alfv\'{e}nic turbulence have an effective mean free path $l_A$ and 
isotropic spatial distribution on scales larger than $l_A$ (Section \ref{sec:turdyn}).
For an oblique shock with small magnetic fluctuations, the inefficient scattering and anisotropic pitch angle distribution can affect the particle energy spectral index
\citep{Kirk89,Be21}. 
For instance, 
as pointed out by 
\citet{NaTa95},
\citet{Kirk89}
considered a particle distribution in the upstream region concentrated at a small $\mu$,
which results in a small probability of transmission and escape and thus 
a flat particle spectrum with the spectral index as low as $1$. 
Given the energy-dependent scattering efficiency and pitch-angle distribution, the particle spectral index 
for an oblique shock may vary in different ranges of particle energies.

In addition to pitch angle scattering and turbulence tangling, 
the {\sy non-resonant mirroring} in compressible MHD turbulence 
\citep{LX21}
can also contribute to the confinement of CRs near the shock. 
The particles   
bouncing with magnetic mirrors generally have a smaller parallel diffusion coefficient than that resulting from gyroresonant scattering.

The diffusion coefficient that depends on the properties of MHD turbulence can vary with the distance from the 
shock. 
The spatial dependence of the diffusion coefficient and its effect on escape of particle in the upstream region 
was considered in, e.g., 
\citet{Fra21}.
It is found that the form of particle spectrum can be modified when a non-constant diffusion coefficient is adopted.


\section{Conclusions}

Shock acceleration is environment-dependent. 
For shocks propagating through inhomogeneous astrophysical media, e.g., the multi-phase ISM, 
it is necessary to consider the effect of upstream density inhomogeneities on preshock magnetic field amplification and shock acceleration.

Oblique shocks are commonly seen in the presence of uniform magnetic field or 
turbulent magnetic fields 
that cause variation of shock obliquity along its face. 
In the limit case of a small $\beta_1$, 
the spectral index of accelerated particles is the same as that of a parallel shock and 
independent of shock obliquity. 
This result is consistent with the findings by, e.g., 
\citet{Drury83,Ostr88}.
As earlier reported by, e.g., 
\citet{NaTa95,Be21},
the obliquity effect on shock acceleration is more significant at a larger shock speed. 
We see both large energy gain and escape probability at a large obliquity. 
The spectral steepening for quasi-perpendicular shocks at large shock speeds has an important 
implication on understanding the steep radio synchrotron spectra of young SNRs
(Fig. \ref{fig: comobs}).

We studied a
new regime of shock acceleration 
in the presence of pre- and post-shock dynamo-amplified magnetic fields. 
Particles following turbulent magnetic fields can have an isotropic spatial distribution and repeatedly cross the shock 
without pitch angle scattering.
The mean energy gain per cycle is larger than that of a parallel shock due to the local obliquities created by turbulent 
magnetic fields, while the escape probability in the downstream region is comparable to that of a parallel shock 
due to the coupling of particles with the fluid facilitated by turbulent magnetic fields. 
As a result, the particle spectrum is much shallower than that of a parallel shock 
and the spectral index has a weak dependence on the shock speed. 
This finding is different from previous finding by 
\citet{Be21}, 
where they argued that the spectral index averaged over random magnetic field orientations is approximately the same 
as that for a parallel shock. 
The shallow particle spectrum that we found in the presence of dynamo-amplified turbulent magnetic fields can be 
important for explaining the flat radio spectra of old SNRs. 
These SNRs are frequently found to have interaction with MCs where the density distribution is characterized by 
small-scale density enhancements.


The diffusion of particles and thus the acceleration time are also affected by turbulent magnetic fields. 
For oblique shocks with sub-Alfv\'{e}nic turbulence,
the anisotropic diffusion of particles is regulated by turbulent magnetic fields, 
and the ratio of perpendicular to parallel diffusion coefficients depends on turbulence $M_A$. 
In super-Alfv\'{e}nic turbulence, the diffusion is isotropic on scales larger than the characteristic scale 
$l_A$ of turbulent magnetic fields.
We found that the acceleration time for a quasi-perpendicular shock with sub-Alfv\'{e}nic turbulence 
and a shock with super-Alfv\'{e}nic turbulence can be significantly shorter than that of a parallel shock 
(Fig. \ref{fig: acctime}).
It means that the acceleration efficiency depends on both shock obliquity and properties of pre- and post-shock 
turbulent magnetic fields.

It is well known that the varying obliquity and turbulence conditions play an important role in particle transport and acceleration  
at heliospheric shocks
\citep{Guo21}. 
Our study on shock acceleration with oblique magnetic fields and turbulent magnetic fields has a wide applicability in diverse astrophysical environments, e.g.,  SN shocks in the turbulent, inhomogeneous, and multi-phase ISM.


\acknowledgments
S.X. acknowledges the support for 
this work provided by NASA through the NASA Hubble Fellowship grant \# HST-HF2-51473.001-A awarded by the Space Telescope Science Institute, which is operated by the Association of Universities for Research in Astronomy, Incorporated, under NASA contract NAS5-26555. 
A.L. acknowledges the support of NASA ATP  AAH7546.
\software{MATLAB \citep{MATLAB:2018}}

\appendix
\section{General expressions of $d$ and $P_\text{esp}$ for an oblique shock}
\label{sec:app}

We first follow the approach in \citet{Ostr88}
to derive the general expression of $d$, which is applicable to the case when 
$\beta_1$ is not much smaller than unity. 
For a subluminal oblique shock, 
it is convenient to introduce the 
de-Hoffmann-Teller (HT) frame, where the electric field vanishes and the particle energy is conserved 
\citep{Drury83}. 
The Lorentz factor of a particle in the local fluid frame is related to that in the HT frame (denoted by $^\prime$) 
by 
\begin{equation}
    \gamma = \Gamma_i \gamma^\prime (1- \beta_i \mu^\prime),
\end{equation}
where $\Gamma_i = 1/\sqrt{1-\beta_i^2}$.
The cosine of the pitch angle in the two frames are related by 
\begin{equation}\label{eq: murela}
    \mu = \frac{\mu^\prime - \beta_i}{ 1 - \beta_i \mu^\prime},  ~~ \mu^\prime =  \frac{\mu + \beta_i}{1+\beta_i \mu}.
\end{equation}

We have the ratio of particle energies after (denoted by ``f") and before (denoted by ``0") a shock encounter as 
\begin{equation}\label{eq: renoec}
   \frac{E_f}{E_0} =  \frac{\gamma_f}{\gamma_0} = \frac{\Gamma_f (1- \beta_f \mu_f^\prime)}{\Gamma_0  (1- \beta_0 \mu_0^\prime)}.
\end{equation}
For a reflected particle, there are 
\begin{equation}
\Gamma_0 = \Gamma_f = \Gamma_1, ~~ \beta_0 = \beta_f = \beta_1 = \frac{V_1}{c}, 
~~ \Gamma_1 = \frac{1}{\sqrt{1-\beta_1^2}}, ~~ \mu_f^\prime = -\mu_0^\prime = -\mu^\prime.
\end{equation}
Therefore, we have 
\begin{equation}
  \Big(\frac{E_f}{E_0}\Big)_r  =  \frac{1+ \beta_1 \mu^\prime}{ 1- \beta_1 \mu^\prime}.
\end{equation}
Then inserting Eq. \eqref{eq: murela} into the above expression and taking $\beta_i = \beta_1$ yields
\begin{equation}
   \Big(\frac{E_f}{E_0}\Big)_r 
   = \Gamma_1^2 (1+ 2 \beta_1 \mu +\beta_1^2).
\end{equation}

For a particle   
transmitted from upstream to downstream regions, there are 
\begin{equation}\label{eq: cotud}
   \Gamma_0 = \Gamma_1, ~\Gamma_f = \Gamma_2, ~ \beta_0 = \beta_1, ~ \beta_f = \beta_2, ~\Gamma_1 = \frac{1}{\sqrt{1-\beta_1^2}},
   ~\Gamma_2 = \frac{1}{\sqrt{1-\beta_2^2}}, ~  \mu_0^\prime = \mu_1^\prime = \mu^\prime, ~ \mu_f^\prime = \mu_2^\prime, 
\end{equation}
Due to the conservation of the magnetic moment, 
$\mu_1^\prime$ and $\mu_2^\prime$ are related by 
\begin{equation}\label{eq: muptcos}
   \mu_1^\prime  = \sqrt{1-b(1-{\mu_2^\prime}^2)},
   ~~~~ \mu_2^\prime  = \sqrt{1-\frac{1}{b}(1-{\mu_1^\prime}^2)} ,
\end{equation}
where $b = B_1 /B_2$.
By combining Eqs. \eqref{eq: renoec}, \eqref{eq: cotud}, and \eqref{eq: muptcos}, we can get 
\begin{equation}
\begin{aligned}
  \Big(\frac{E_f}{E_0}\Big)_{12} & = \frac{\Gamma_2 (1- \beta_2 \mu_2^\prime)}{\Gamma_1  (1- \beta_1 \mu_1^\prime)} \\
                         & =  \frac{\Gamma_2 \Big(1- \beta_2 \sqrt{1-\frac{1}{b}(1-{\mu^\prime}^2)} \Big)}{\Gamma_1  (1- \beta_1 \mu^\prime)} .
\end{aligned}
\end{equation}
Then by inserting Eq. \eqref{eq: murela} into the above expression and using $\beta_i = \beta_1$, we find 
\begin{equation}
\begin{aligned}
  \Big(\frac{E_f}{E_0}\Big)_{12} 
                         =  \Gamma_2 \Gamma_1  \Bigg(1+\beta_1 \mu- \beta_2  
                         \sqrt{(1+\beta_1 \mu)^2-\frac{1}{b}  \frac{1-\mu^2}{ \Gamma_1^2}} \Bigg) .
\end{aligned}
\end{equation}

In the case of transmission from downstream to upstream regions, there are 
\begin{equation}\label{eq: ducosl}
   \Gamma_0 = \Gamma_2, ~\Gamma_f = \Gamma_1, ~ \beta_0 = \beta_2, ~ \beta_f = \beta_1, ~\Gamma_1 = \frac{1}{\sqrt{1-\beta_1^2}},
   ~\Gamma_2 = \frac{1}{\sqrt{1-\beta_2^2}}, ~  \mu_0^\prime = \mu_2^\prime = \mu^\prime, ~ \mu_f^\prime = \mu_1^\prime, 
\end{equation}
and $\mu_1^\prime$ and $\mu_2^\prime$ are related by
\begin{equation}\label{eq: dupmr}
   \mu_1^\prime  = -\sqrt{1-b(1-{\mu_2^\prime}^2)}.
\end{equation}
Combining Eqs. \eqref{eq: renoec}, \eqref{eq: ducosl}, and \eqref{eq: dupmr} leads to 
\begin{equation}
\begin{aligned}
   \Big(\frac{E_f}{E_0}\Big)_{21} 
                          &= \frac{\Gamma_1 (1- \beta_1 \mu_1^\prime)}{\Gamma_2  (1- \beta_2 \mu_2^\prime)} \\
                          &= \frac{\Gamma_1 (1+ \beta_1 \sqrt{1-b(1-{\mu^\prime}^2)})}{\Gamma_2  (1- \beta_2 \mu^\prime)} .
\end{aligned}
\end{equation}
By inserting Eq. \eqref{eq: murela} into the above expression and using $\beta_i = \beta_2$, we obtain 
\begin{equation}
\begin{aligned}
   \Big(\frac{E_f}{E_0}\Big)_{21} 
                            =  \Gamma_1 \Gamma_2  \Bigg(1+\beta_2\mu+ \beta_1 \sqrt{(1+\beta_2\mu)^2-b \frac{1-\mu^2}{\Gamma_2^2}}\Bigg) .
\end{aligned}
\end{equation}
The above results are consistent with those in \citet{Ostr88}. 
We note that in the limit of a parallel shock, Eq. \eqref{eq: renoec} becomes 
\begin{equation}
    \frac{E_f}{E_0} = 1 + \mu \frac{3U_1}{4c}. 
\end{equation}

The range of $\mu^\prime$ is  
\begin{equation}
   0<\mu^\prime <\sqrt{1-b}
\end{equation}
for reflected particles,
\begin{equation}
   \sqrt{1-b}<\mu^\prime <1
\end{equation}
for transmitted particles from upstream to downstream regions, and  
\begin{equation}
   -1< \mu^\prime <0
\end{equation} 
for transmitted particles from downstream to upstream regions. Using the relation in Eq. \eqref{eq: murela}, 
the range of $\mu$ is 
\begin{equation}
   -\beta_1<\mu <\frac{\sqrt{1-b}-\beta_1}{1-\beta_1\sqrt{1-b}}
\end{equation}
for reflected particles, 
\begin{equation}
    \frac{\sqrt{1-b}-\beta_1}{1-\beta_1\sqrt{1-b}}    <\mu < 1
\end{equation} 
for transmitted particles from upstream to downstream regions, and 
\begin{equation}
  -1< \mu < - \beta_2
\end{equation}
for transmitted particles from downstream to upstream regions.

The mean fractional energy gain during a shock encounter is given by 
\citep{Ostr88},
\begin{equation}
    d_x = \frac{\int V_{rel} \Big(\frac{E_f}{E_0}-1\Big) d\mu}{S}, 
\end{equation}
where ``$x$" is ``r", ``12", or ``21",
the weight function is 
\begin{equation}
   V_{rel,1} = \frac{\mu \cos\alpha_1 + \frac{U_1}{c}}{1-\mu \cos\alpha_1 \frac{U_1}{c}}, ~~
   V_{rel,2} = \frac{\mu\cos \alpha_2 - \frac{U_2}{c} }{1 + \mu \cos\alpha_2 \frac{U_2}{c}}
\end{equation} 
in the upstream and downstream regions, respectively, and 
\begin{equation}
   S = \int V_{rel} d\mu.
\end{equation} 
With the probabilities of reflection and transmission from upstream to downstream given by 
\citep{Ostr88},
\begin{equation}
   P_r = \frac{S_r}{S_{12}+S_r}, ~~ P_{12} = \frac{S_{12}}{S_{12}+S_r},
\end{equation}
the general form of the mean fractional energy gain per cycle is 
\citep{Ostr88},
\begin{equation}
   d = P_{12} (d_{12}+d_{21}) + P_r d_r. 
\end{equation}

The probability of a particle that is transmitted downstream and does not return to the shock is 
\begin{equation}\label{eq: pespapp}
    P_\text{esp} = P_{12} \frac{F_2^\text{esp}}{F_2}, 
\end{equation} 
where $F_2^\text{esp}$ the flux of escaping particles, and 
$F_2$ is the flux entering into the downstream region. To the approximation order $O(\beta_1)$, 
there are 
\citep{Drury83}, 
\begin{equation} 
     F_2^\text{esp} = n_2V_2 \cos \alpha_2, ~~~~
     F_2 = \frac{n_2}{4} c \cos\alpha_2, ~~~~
     \frac{F_2^\text{esp}}{F_2} = 4 \beta_2,
\end{equation}
where $n_2$ is the number density of particles in the downstream fluid frame.  To obtain the general expression of $P_\text{esp}$, 
we use 
\citep{NaTa95} 
\begin{equation}\label{eq: fnetfgen}
    F_2^\text{esp} =\Gamma_2 n_2 V_2 \cos \alpha_2, 
    ~~~~ F_2 = \frac{\Gamma_2 n_2}{4} c \frac{1}{\Gamma_2^4 (1-\beta_2)^2} \cos\alpha_2, 
    ~~~~\frac{F_2^\text{esp}}{F_2} = \frac{4\beta_2}{(1+\beta_2)^2},
\end{equation} 
where $\Gamma_2 n_2$ is the number density of particles in the HT frame.
The general expression of $F_2^\text{esp}/F_2$ is smaller than the approximate one 
when $\beta_2$ is not very small.

\begin{figure*}[ht]
\centering
\subfigure[]{
   \includegraphics[width=8.7cm]{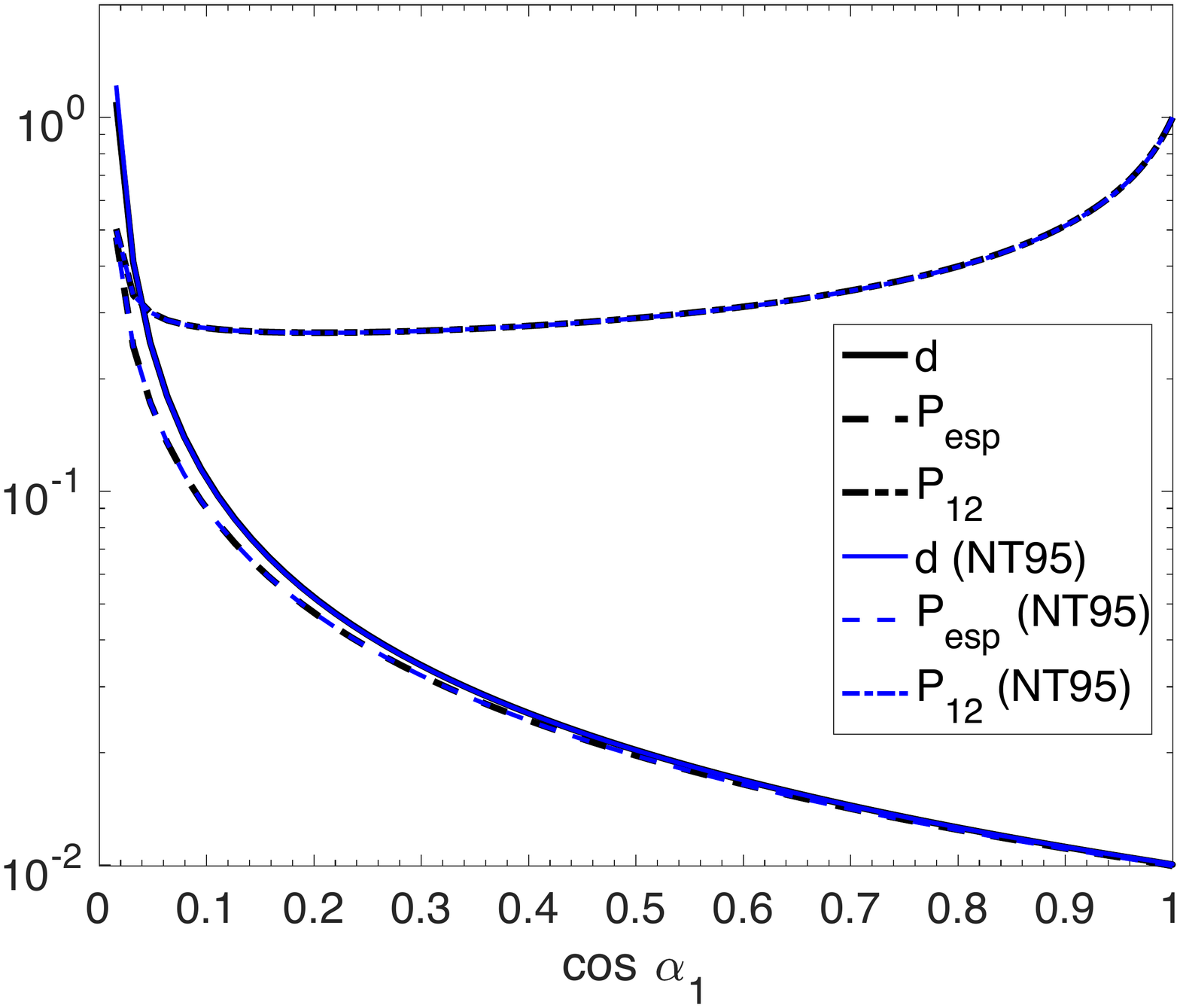}\label{fig: comdpesp}}
\subfigure[]{
   \includegraphics[width=8.7cm]{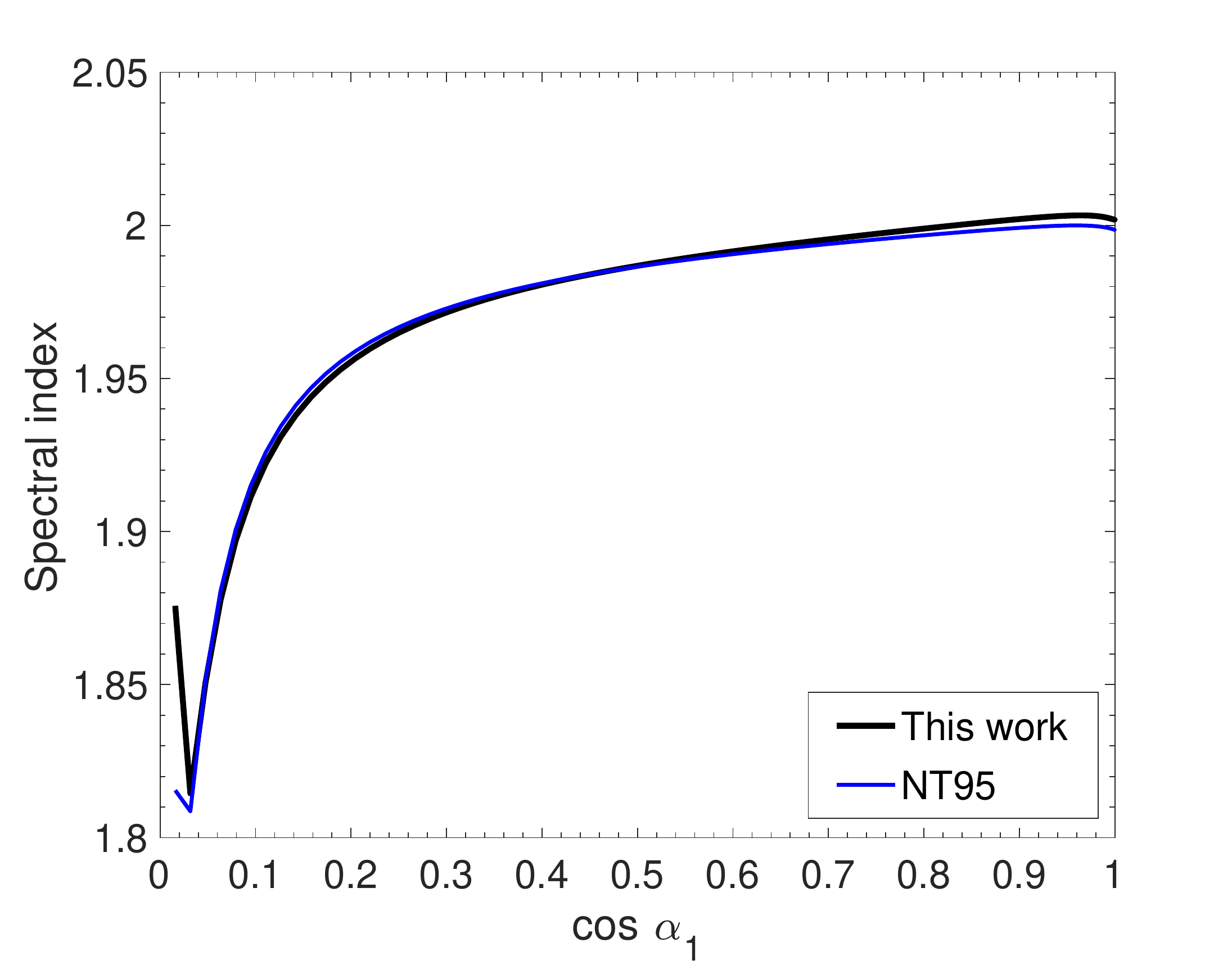}\label{fig: comspeind}}
\caption{Comparison between the results calculated using our general expressions and those from 
\citet{NaTa95} (NT95) for $U_1=0.01c$.}
\label{fig: comshoiso}
\end{figure*}

\begin{figure*}[ht]
\centering
\subfigure[]{
   \includegraphics[width=8.7cm]{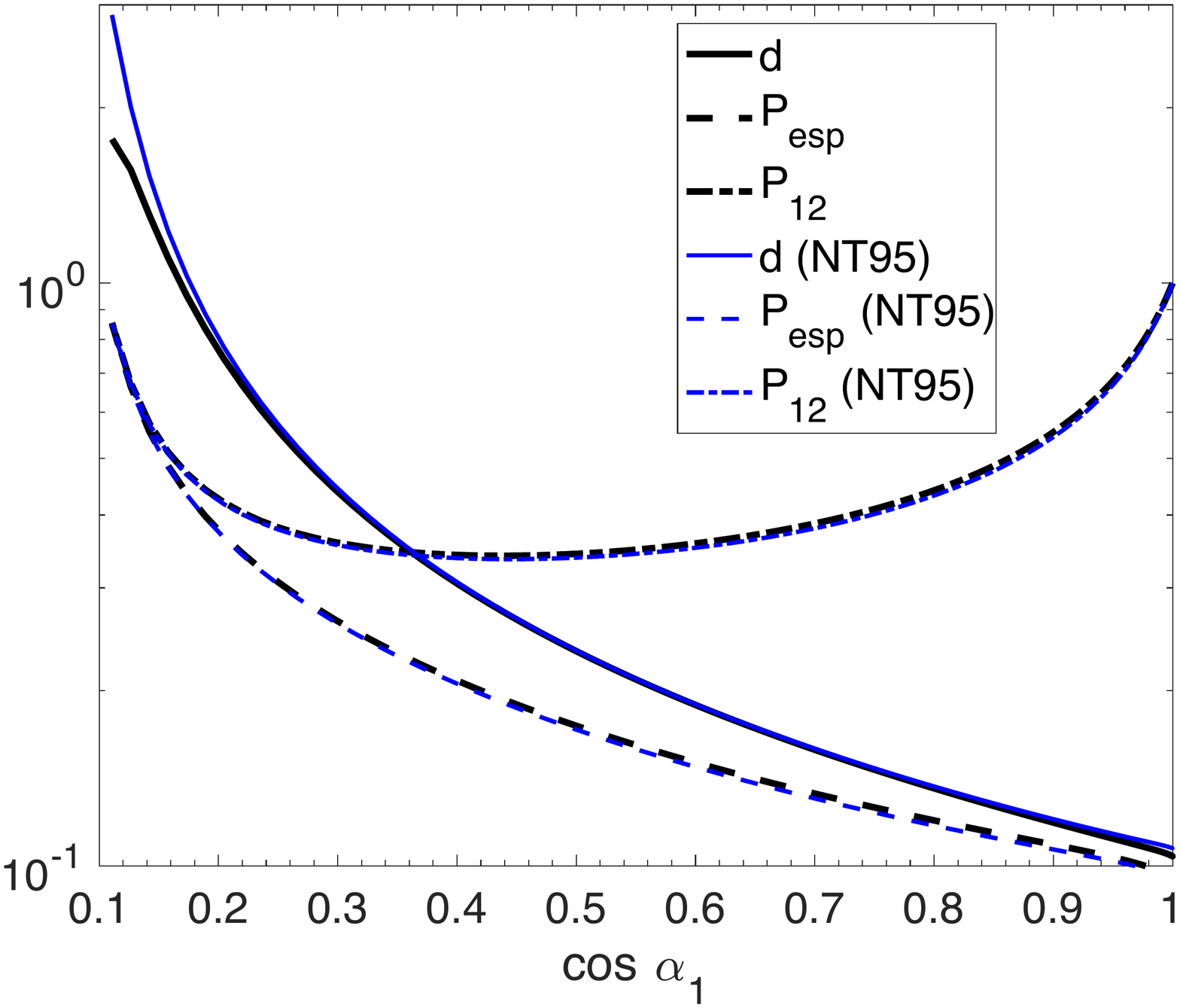}\label{fig: comdpesp2}}
\subfigure[]{
   \includegraphics[width=8.7cm]{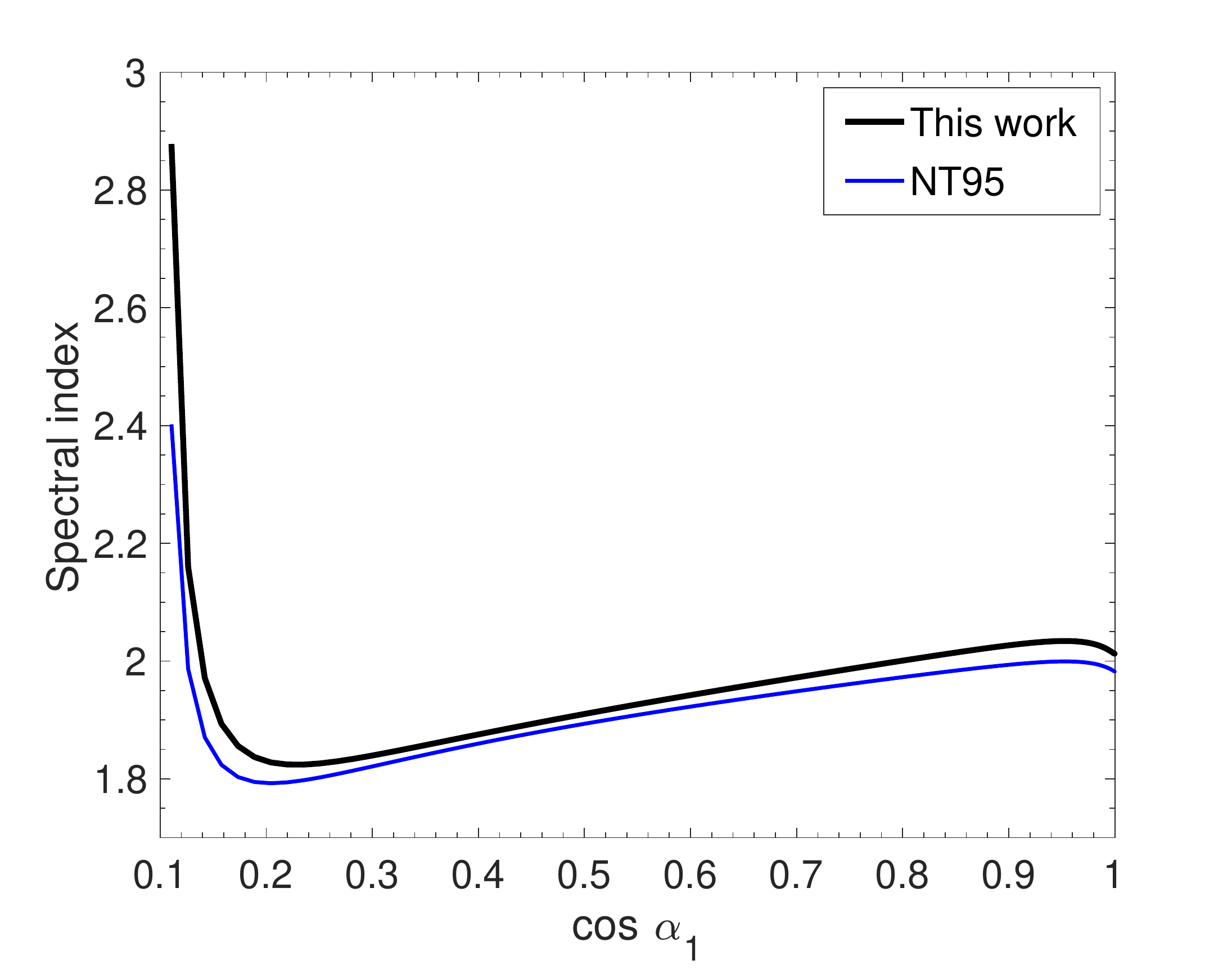}\label{fig: comspeind2}}
\caption{Same as Fig. \ref{fig: comshoiso} but for $U_1=0.1c$.}
\label{fig: comshoiso2}
\end{figure*}

In Fig. \ref{fig: comshoiso} and Fig. \ref{fig: comshoiso2}, 
we compare the results calculated using the above general expressions 
with those calculated using the expressions in  
\citet{NaTa95}, 
where they also considered a general case but adopted a different approach. 
We see a good agreement.  
 
\bibliographystyle{aasjournal}
\bibliography{xu}

\end{document}